\documentclass[aps,twocolumn,prl,nofootinbib,longbibliography,
amsmath,amssymb,amsfonts,superscriptaddress,notitlepage]{revtex4-2}
            
\usepackage[varg]{txfonts}

\usepackage[T1]{fontenc}
\usepackage[utf8]{inputenc}

\usepackage[bookmarksnumbered,pdfpagelabels=true,plainpages=false,colorlinks=true,linkcolor=blue,citecolor=red,urlcolor=blue]{hyperref}
\usepackage{amsmath}
\usepackage{color}
\usepackage{float}
\usepackage{graphicx}
\usepackage[percent]{overpic}
\usepackage{mathrsfs}
\usepackage{bm}
\usepackage{braket}
\usepackage{mathtools}
\usepackage{ulem}

\usepackage{tikz}

\definecolor{cL}{RGB}{59, 83, 140}
\definecolor{cM}{RGB}{33, 145, 141}
\definecolor{cH}{RGB}{95, 202, 98}
\definecolor{cG}{RGB}{204,204,204}

\renewcommand{\vec}[1]{\boldsymbol{#1}}

\def\be{\begin{equation}}
\def\ee{\end{equation}}

\def\singlei{\tikz[baseline=.1ex]{\begin{scope}[scale=0.3, yshift = 0.4cm]
\draw (-0.5,0)--(0.5,0);
	\draw (1, 0.866)--(0.5,0);
	\draw (1,-0.866)--(0.5,0);
	\draw (-0.5,0)--(-1, 0.866);
	\draw (-0.5,0)--(-1,-0.866);

	\draw[fill=black] (-0.5,0) circle (0.15cm) node[left] {$i$};
	\draw[fill=white] ( 0.5,0) circle (0.15cm) node[right] {$j$};
	\draw[fill=white] (   1, 0.866) circle (0.15cm);
	\draw[fill=white] (   1,-0.866) circle (0.15cm);
	\draw[fill=white] (  -1, 0.866) circle (0.15cm);
	\draw[fill=white] (  -1,-0.866) circle (0.15cm);	
	\end{scope}
	}
}

\def\singlej{\tikz[baseline=.1ex]{\begin{scope}[scale=0.3, yshift = 0.4cm]
\draw (-0.5,0)--(0.5,0);
	\draw (1, 0.866)--(0.5,0);
	\draw (1,-0.866)--(0.5,0);
	\draw (-0.5,0)--(-1, 0.866);
	\draw (-0.5,0)--(-1,-0.866);

	\draw[fill=white] (-0.5,0) circle (0.15cm) node[left] {$i$};
	\draw[fill=black] ( 0.5,0) circle (0.15cm) node[right] {$j$};
	\draw[fill=white] (   1, 0.866) circle (0.15cm); 
	\draw[fill=white] (   1,-0.866) circle (0.15cm);
	\draw[fill=white] (  -1, 0.866) circle (0.15cm);
	\draw[fill=white] (  -1,-0.866) circle (0.15cm);	
	\end{scope}
	}
}

\def\doubleij{\tikz[baseline=.1ex]{\begin{scope}[scale=0.3, yshift = 0.4cm]
\draw (-0.5,0)--(0.5,0);
	\draw (1, 0.866)--(0.5,0);
	\draw (1,-0.866)--(0.5,0);
	\draw (-0.5,0)--(-1, 0.866);
	\draw (-0.5,0)--(-1,-0.866);

	\draw[fill=black] (-0.5,0) circle (0.15cm) node[left] {$i$};
	\draw[fill=black] ( 0.5,0) circle (0.15cm) node[right] {$j$};
	\draw[fill=white] (   1, 0.866) circle (0.15cm);
	\draw[fill=white] (   1,-0.866) circle (0.15cm);
	\draw[fill=white] (  -1, 0.866) circle (0.15cm);
	\draw[fill=white] (  -1,-0.866) circle (0.15cm);	
	\end{scope}
	}
}

\definecolor{cL}{RGB}{59, 83, 140}
\definecolor{cM}{RGB}{33, 145, 141}
\definecolor{cH}{RGB}{95, 202, 98}
\definecolor{cG}{RGB}{204,204,204}

\begin{document}

\title{Spurious Symmetry Enhancement and Interaction-Induced Topology in Magnons}

\author{Matthias Gohlke}

\affiliation{Theory of Quantum Matter Unit, Okinawa Institute of Science and Technology Graduate University, Onna-son, Okinawa 904-0495, Japan}

\author{Alberto Corticelli}
\affiliation{Max Planck Institute for the Physics of Complex Systems, N\"{o}thnitzer Str. 38, 01187 Dresden, Germany}

\author{Roderich Moessner}
\affiliation{Max Planck Institute for the Physics of Complex Systems, N\"{o}thnitzer Str. 38, 01187 Dresden, Germany}

\author{Paul~A.~McClarty}
\affiliation{Max Planck Institute for the Physics of Complex Systems, N\"{o}thnitzer Str. 38, 01187 Dresden, Germany}

\author{Alexander Mook}

\affiliation{Institute of Physics, Johannes Gutenberg University Mainz, 55128 Mainz, Germany}

\begin{abstract}
Linear spin wave theory (LSWT) is the standard technique to compute the spectra of magnetic excitations in quantum materials. In this paper, we show that LSWT, even under ordinary circumstances, may fail to implement the symmetries of the underlying ordered magnetic Hamiltonian leading to spurious degeneracies. In common with pseudo-Goldstone modes in cases of quantum order-by-disorder these degeneracies tend to be lifted by magnon-magnon interactions. We show how, instead, the correct symmetries may be restored at the level of LSWT. In the process we give examples, supported by nonperturbative matrix product based time evolution calculations, where symmetries dictate that there should be a topological magnon gap but where LSWT fails to open up this gap. We also comment on possible spin split magnons in MnF$_2$ and similar rutiles by analogy to recently proposed altermagnets. 
\end{abstract}

\maketitle

From N\'{e}el order in the mid 20th century to skyrmion phases in the 21st, magnetically ordered materials have been a constant source of insights into the collective behavior of matter. The coherent spin wave excitations, or magnons, about these magnetic textures provide invaluable information about magnetic structures and couplings. They are also interesting in their own right: as a window into many-body interactions and quasiparticle breakdown \cite{zhitomirskychernyshev2013}, as a platform for investigating band topology \cite{malki2020topological, bonbien2021topological,mcclarty2022ar}, and as an essential ingredient in the functioning of many spintronics devices \cite{chumak2015magnon}. 


One of the most useful theoretical tools at our disposal to understand magnons is an expansion in powers of inverse spin $S$ based on the Holstein-Primakoff bosonization of quantum spins \cite{holsteinprimakoff1940}. The single particle spectrum arising from spin wave theory to quadratic order (called linear, or non-interacting, spin wave theory) is often used with great success to constrain magnetic couplings from experimental data. This theory is known to fail qualitatively in cases where coupling between single and multi-particle states becomes important for example in highly frustrated magnets and non-collinear spin textures such as the famous triangular lattice antiferromagnet \cite{starykh2006,chernyshevzhitomirsky2009,zhitomirskychernyshev2013}, and close to quantum phase transitions \cite{sachdev2011quantum}. 

Another, more subtle way, in which linear spin wave theory (LSWT) can fail qualitatively is called order-by-disorder \cite{villain1980order,shender1982antiferromagnetic,henley1989,chalker1992} where spurious ground states and symmetry enhancement exist at the semi-classical level that are lifted by fluctuations. 
In some instances of quantum order-by-disorder, a spurious continuous symmetry forces the presence of a pseudo-Goldstone mode in LSWT where none should be present \cite{rau2018}. In this paper, we focus on a related instance of this physics where, instead of failing to capture degeneracy breaking in the ground state, the LSWT instead does not fully capture symmetries that affect degeneracies higher up in the excitation spectrum \cite{mook2021}. 
With growing interest in magnon band topology \cite{Katsura2010, Onose2010, Ideue2012, zhang2013, Hoogdalem2013, Shindou13, Shindou13b, Shindou14, Mook2014, mook2014edge, owerre2016first,chisnell2015topological,Chen2018,fransson2016, Mook2016Weyl, Xu2016, McClarty2018, Chen2018CrI3, yuan2020dirac, Mook2021hinge, Scheie2022}, there is additional impetus to understand how to implement symmetries correctly in LSWT as these provide important constraints on the possible topological band structures that can arise \cite{mook2021,corticelli2022,corticelli2022b}. 

\begin{figure}
    \centering
    \includegraphics[width=\linewidth,trim={0 1.5cm 0 0},clip]{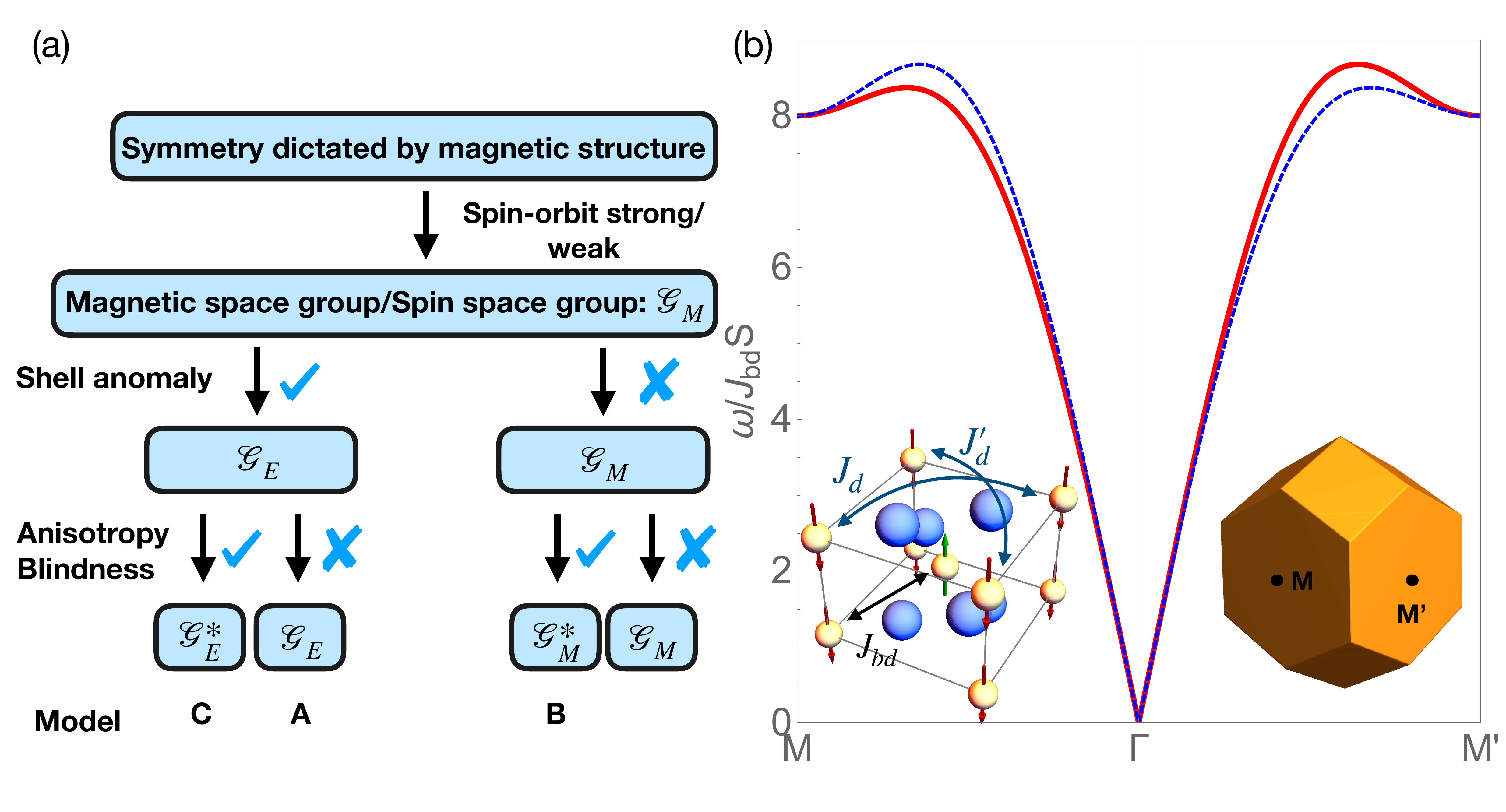}
    \caption{(a) Flow chart indicating the actual and spurious symmetries that may arise in LSWT. The parent symmetry group $\mathcal{G}_M$ constrained by the lattice symmetries, magnetic structure and the nature of the exchange may be enhanced via a \textit{shell anomaly} and/or \textit{anisotropy blindness} to new symmetry groups denoted $\mathcal{G}^{(*)}_E$ or $\mathcal{G}^{*}_M$. (b) Spin wave spectrum for a model A exhibiting a shell anomaly in the form  of a double degeneracy in the absence of further neighbor terms. The spectrum shown (for the lattice model in the inset) has $J_d \neq J'_d$.}
    \label{fig:squareexch}
\end{figure}

{\it Overview.}
In many-body physics, there is a large body of work on cases where at long wavelengths and low energies there are enhanced symmetries (e.g. \cite{lin1998,james2018}). In contrast, the goals of this paper are to spell out ways in which spin wave theory can lead to spurious degeneracies in excitations across the Brillouin zone and to supply a simple, general way to resolve them.

The cases we consider fall into two classes [see Fig.~\ref{fig:squareexch}(a)]. 
The first class is where the lattice symmetries are not manifest for exchange couplings between moments out to $n$th nearest neighbors but where the symmetries do manifest for longer-range couplings. This, we call the {\it shell anomaly}. Such a situation may be completely physical and, far from being confined to spin wave theory, it may arise in general tight-binding models. The second class is more subtle: where LSWT does not capture certain kinds of exchange anisotropy or {\it anisotropy blindness}. Then, LSWT fails to produce the correct magnon spectrum at a qualitative level and spurious symmetry-protected topological magnon degeneracies occur. We show that degeneracy breaking occurs by carrying out DMRG plus matrix product operator time evolution (DMRG+tMPO) \cite{white_density_1992,phien_infinite_2012,zaletel_timeevolving_2015} to resolve band splittings nonperturbatively. While the most straightforward LSWT does not capture the symmetries of the magnetic Hamiltonian, one may show that the symmetry breaking terms, treated perturbatively, lead to effective magnon hopping terms that do resolve spurious degeneracies. This fact leads us to propose a general solution to the problem by including all symmetry-allowed exchange couplings out to some shell.


The basic mechanism of both classes are explained in more detail below. Figure~\ref{fig:squareexch}(a) is a schematic overview of the paper from a symmetry perspective. If the symmetry group dictated by the lattice, the magnetic ground state, and the presence or absence of exchange anomalies is $\mathcal{G}_M$, the symmetries may be enhanced by the shell anomaly (model A), anisotropy blindness (model B) or both (model C) leading to new symmetry groups. The models A, B, and C are discussed below and serve as worked examples that make contact with material classes such as altermagnets \cite{Smejkal2021altermagnets,Smejal2022}, chiral magnets \cite{Garst2017}, and van der Waals magnets \cite{Burch2018}.

{\it Shell Anomaly and Connection to Altermagnetism.} We start with an example  that illustrates the shell anomaly (model A). Figure~\ref{fig:squareexch}(b) shows the crystal and magnetic structure of MnF$_2$. The symmetries ensure that there is a single nearest neighbor coupling $J_{\rm bd}$ on all the bonds joining the two magnetic sublattices in a primitive cell. In MnF$_2$ this is antiferromagnetic and, for this coupling alone, the model is identical to the simple body-centred tetragonal antiferromagnet with a double (spin) degeneracy in the magnon spectrum. However, further neighbor exchange will, in general, lift the double degeneracy [see Supplemental Material (SM) \cite{SM} for details]. Specifically, if the further neighbor couplings $J_d$ and $J'_d$ are unequal, as is allowed by the symmetry of the lattice including the fluoride ion positions, the magnon bands are non-degenerate [cf.~Fig.~\ref{fig:squareexch}(b)]. In this instance, not only the linear theory but in fact the exact spin wave theory has an enhanced symmetry at the nearest neighbor level that is lifted by further neighbor couplings. This splitting is identical to the zero spin-orbit coupled electronic d-wave spin splitting reported in Refs.~\onlinecite{Naka2019,Smejkal2021altermagnets,Smejal2022} that goes under the name {\it altermagnetism}.

In the language of group theory introduced above, the nearest neighbor model has a spurious sublattice symmetry present in $\mathcal{G}_E$ and absent in the full symmetry group $\mathcal{G}_M$. A shell anomaly may occur in materials where the exchange couplings are strictly short range.  It is problematic when the exchange couplings in the material break down these symmetries but where this fact is overlooked by the choice of model. 

We note that this shell anomaly can be entirely physical: in the case of MnF$_2$ inelastic neutron data reveals no degeneracy breaking and, thus, the shell anomaly is active in this material to within instrumental resolution \cite{Nikotin_1969,FeF2Magnon,SM}.

{\it Anisotropy Blindness.} We now describe the origin of anisotropy blindness. Consider the bilinear magnetic couplings $J_{ij}^{\alpha\beta}$ between moments labelled by $i$ and $j$ with respective components $\alpha$ and $\beta$ in the quantization frame. Since LSWT is formulated in terms of the transverse spin fluctuations, transverse-longitudinal components $J_{ij}^{z\pm}$ do not enter into the theory. But, for example, the magnetic Hamiltonian with these couplings may have lower symmetry than the 
Hamiltonian without them. In such a situation, one can generally expect that LSWT will fail to capture certain instances of degeneracy breaking in the magnon spectrum. 

A solution to this problem may be simply stated: the transverse-longitudinal components re-enter the transverse components of the dynamical structure factor to higher order in perturbation theory. More precisely, the  $J_{ij}^{z\pm}$ lead to cubic vertices. Then, bubble diagrams with a pair of such vertices dress the single magnon propagator restoring the correct symmetry of the magnon spectrum. While simple in principle, this is burdensome in practice. 
Taking model B as an example, we show how the correct symmetries can be implemented instead already on the level of LSWT within a real-space perturbation theory, as verified by the nonperturbative DMRG+tMPO.

\begin{figure*}[tb]
    \centering
    \includegraphics[width=1.0\linewidth]{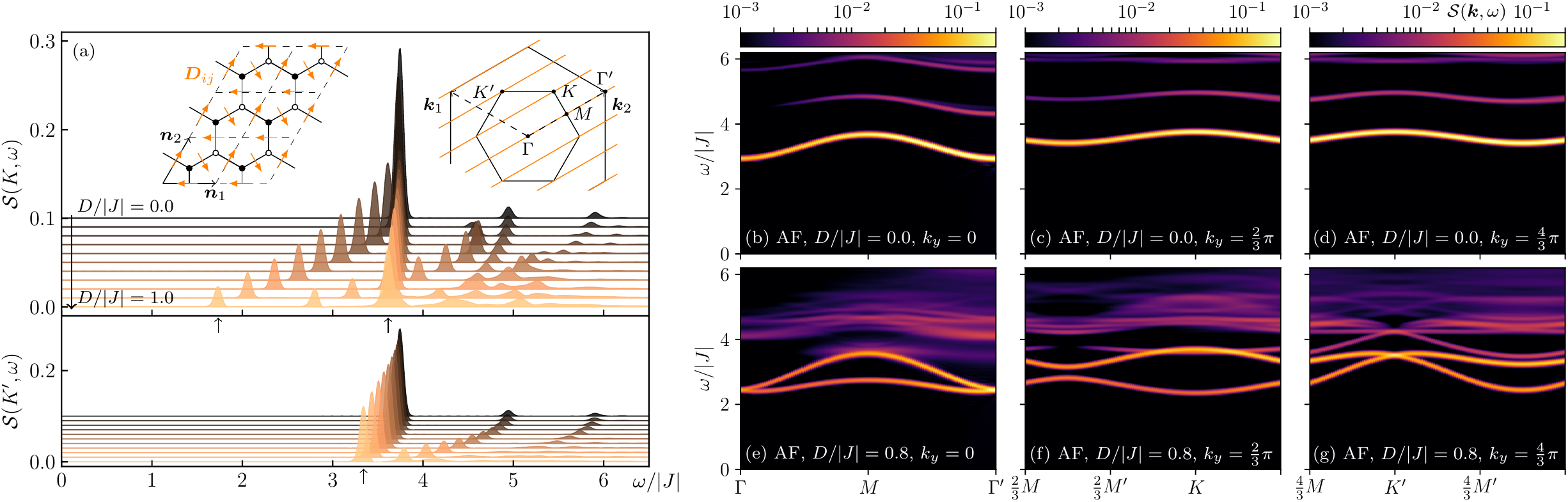}
    \caption{
               Dynamical spin-structure factor $\mathcal{S}(\vec{k},\omega)$ for the spin-$1/2$ honeycomb lattice antiferromagnet, $J_z/|J|=2.4$, and DMI obtained by numerically time-evolving a matrix product state~\cite{zaletel_timeevolving_2015}.
        (a) Line plots at the high-symmetry points $K$ (top) and $K'$ (bottom) for increasing DMI from $D/|J|=0$ to $D/|J|=1$ illustrate the splitting of the 
        spin-wave bands at $K$ while the splitting is absent at $K'$. 
        Magnon bands are highlighted by arrows.
        Insets show the lattice and the corresponding momenta lines determined by the cylindrical geometry with six sites circumference.
        (b-g) Representative color plots along aforementioned momenta cuts for zero DMI (top) and $D/|J| = 0.8$ (bottom). The magnon bands (bright yellow features) split across the entire Brillouin zone apart from the $\Gamma$ and $K'$ points that feature Dirac cones.
        }
    \label{fig:MPS_dyn_AF}
\end{figure*}

\paragraph{Model B1: Honeycomb Lattice Antiferromagnet.}

As a first illustration of the problems faced by LSWT through the omission of longitudinal-transverse couplings (anisotropy blindness), we consider the honeycomb lattice spin-$1/2$ model with nearest neighbor Heisenberg coupling and interfacial Dzyaloshinskii-Moriya interaction (DMI), see inset in Fig.~\ref{fig:MPS_dyn_AF}(a). The Hamiltonian is
\begin{align}
    H 
    &= 
    \frac{1}{2} 
    \sum_{\langle i j \rangle} \left[ 
    J_z S_i^z S_j^z + J \left( S_i^x S_j^x + S_i^y S_j^y \right)
    +
    \vec{D}_{ij} \cdot \vec{S}_i \times \vec{S}_{j}
    \right],
    \label{eq:HamiltonianAFM}
\end{align}
with $J_z > J > 0$ being antiferromagnetic and $\vec{D}_{ij} = D \hat{\vec{z}} \times \hat{\vec{e}}_{ij}$; $\hat{\vec{e}}_{ij}$ is a unit vector along bond direction and $\hat{\vec{z}}$ along the lattice normal.
For $J_z \gg J$ the strong easy-axis Ising anisotropy stabilizes the N\'{e}el ground state. The full model with group $\mathcal{G}_M$ has only discrete symmetries whereas the model without DMI has  symmetry group $\mathcal{G}_M^*$ with a U(1) symmetry. 

Expanding in fluctuations (sublattice-dependent bosons $a$ and $b$),
one obtains the harmonic Hamiltonian in $\vec{k}$ space
$
    H_2(J,J_z) 
    = 
    \frac{1}{2} \sum_{\vec{k}} \vec{\psi}^\dagger_{\vec{k}} H_{\vec{k}} \vec{\psi}_{\vec{k}}
$
featuring a block-diagonal kernel
$
    H_{\vec{k}} = \mathrm{diag}( h_{\vec{k}}, h_{-\vec{k}} )
$
in the basis
$
    \vec{\psi}^\dagger_{\vec{k}} = ( a^\dagger_{\vec{k}}, b_{-\vec{k}}, b^\dagger_{\vec{k}}, a_{-\vec{k}} )
$
with
\begin{align}
    h_{\vec{k}} 
    = 
    \frac{1}{2}
    \begin{pmatrix}
    	3 J_z & - J \gamma_{\vec{k}} \\
    	- J \gamma_{-\vec{k}} & 3 J_z
    \end{pmatrix}, \quad
	\gamma_{\vec{k}} = \sum_{n=0}^2 \mathrm{e}^{\mathrm{i} \vec{k} \cdot \vec{\delta}_n}.
\end{align}
The nearest-neighbor bonds are $\vec{\delta}_n = (\cos\phi_n,\sin\phi_n)$ with $\phi_n=2\pi n/3 + \pi/2$. After diagonalization, we find the normal mode magnon energies
$
    \varepsilon_{\vec{k},\sigma} 
    = 
    \frac{1}{2}
    \sqrt{ (3 J_z)^2 - J^2 | \gamma_{\vec{k}} |^2 },
$
which are two-fold spin-degenerate over the entire Brillouin zone ($\sigma = \uparrow, \downarrow$). This degeneracy is a result of the spurious U(1) and $PT$ symmetries in the LSWT. They appear because the harmonic theory is blind to the symmetry-breaking DMI which enters, to lowest order, to cubic order in the bosons. For a qualitative discussion of magnon-magnon interactions and their influence on the magnon spectrum see SM \cite{SM}.  

We explore the effects of the DMI by carrying out a real space perturbation theory \cite{zhitomirsky_2015},
taking the Ising interaction as the unperturbed Hamiltonian and all other interactions as perturbations.
Processes lifting the band degeneracy are found to second-order in the DMI-induced perturbation $V_D$ via virtual two-spin-flip states: 
\begin{align}
	\frac{-2}{J_z} \left\langle \singlej \right| 
	V_D 
	\left| \doubleij \right\rangle 
	\left\langle \doubleij \right| 
	V_D 
	\left| \singlei \right\rangle 
	& \propto
	\frac{D^2}{J_z} \mathrm{e}^{-2\mathrm{i} \varphi_{ij}},
	\nonumber
\end{align} 
where $\tan \varphi_{ij} = D_{ij}^y/D_{ij}^x$. The states depict the pattern of spin flips generated by $V_D$. White (black) circles indicate the ground state (spin flips).
Such a coupling mimicks the bond-dependent symmetric off-diagonal exchange interaction that breaks spin conservation:
$
    \mathrm{e}^{-2\mathrm{i} \varphi_{ij}} \frac{D^2}{J_z} S^+_i S^+_j
$.
Thus, by taking these terms to replace the DMI in Eq.~\eqref{eq:HamiltonianAFM}, we account for the qualitative effects of DMI. We consider the amended Hamiltonian
\begin{align}
    H'
    = 
    \sum_{i \in \text{A}} \sum_{j = 0}^2 &\left[ 
    J_z S_i^z S_{i+\vec{\delta}_j}^z + \frac{J}{2} \left( S_i^+ S_{i+\vec{\delta}_j}^- + S_i^- S_{i+\vec{\delta}_j}^+ \right)
    \right. \nonumber \\
	&+\left. \frac{J_{++}'}{2} \left( \mathrm{e}^{\mathrm{i} \vartheta_{\vec{\delta}_j}} S_i^+ S_{i+\vec{\delta}_j}^+ + \mathrm{e}^{-\mathrm{i} \vartheta_{\vec{\delta}_j}} S_i^- S_{i+\vec{\delta}_j}^- \right)  
    \right],
    \label{eq:HamiltonianAFM2}
\end{align}
where $|J_{++}'| \propto D^2/J_z$. The $i$-sum runs over all sites of the A sublattice (spin-up) and the phases along the nearest-neighbor bonds read $\vartheta_{\vec{\delta}_n} = 2\pi n /3$.
A LSWT of $H'$ yields a bilinear Hamiltonian $H'_2(J,J_z,D)$ that no longer features a block-diagonal kernel. As a result, the degeneracy of the magnon modes is lifted throughout the Brillouin zone except for the $\Gamma$ and the $K'$ point, which feature magnon Dirac cones, in agreement with Refs.~\onlinecite{Matsumoto2020, Neumann2022}; see the SM \cite{SM} for details.

The qualitative predictions of the modified LSWT are borne out by a fully nonperturbative calculation on the original model, Eq.~\eqref{eq:HamiltonianAFM}. Figure~\ref{fig:MPS_dyn_AF} shows the dynamical spin-structure factor obtained from DMRG+tMPO; 
see SM for technical details \cite{SM}. Constant momentum slices show the progressive splitting of the bands at $K$ as a function of the DMI coupling [Fig.~\ref{fig:MPS_dyn_AF}(a)]; in contrast, the magnon bands stay degenerate at $K'$. As predicted, the DMI lifts the double degeneracy of the single magnon levels almost everywhere in the zone with exceptions at $\Gamma$ and $K'$, where we find \textit{interaction-induced Dirac magnons} [Figs.~\ref{fig:MPS_dyn_AF}(b-g)].

\begin{figure}
    \centering
    \includegraphics[width=0.95\linewidth]{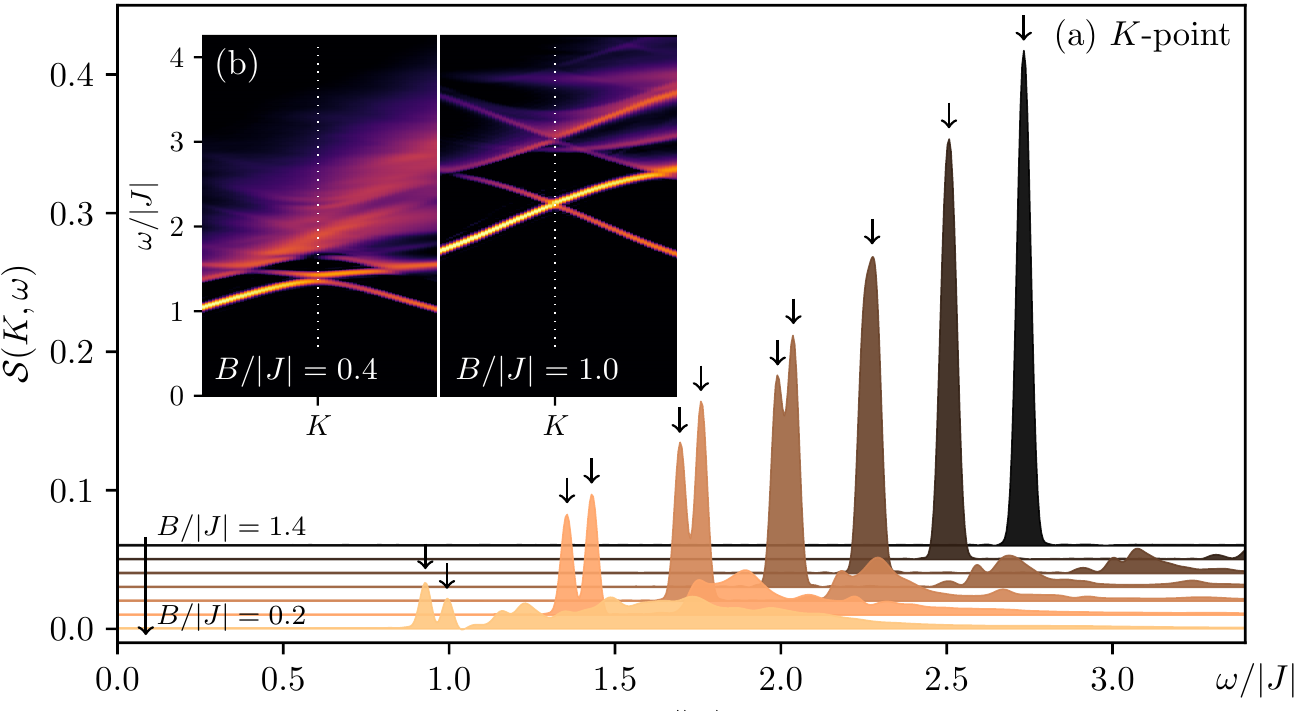}
    \caption{
        Numerically obtained dynamical spin-structure factors $\mathcal{S}(\vec{k},\omega)$ for the spin-$1/2$ honeycomb lattice ferromagnet with magnetic field, $B$, and DMI, $D/|J|=0.6$.
        (a) Line plots at the high-symmetry point $K$ illustrating the splitting of the Dirac-like mode in between the two magnon bands (highlighted by arrows).
        (b) Representative color plots of $\mathcal{S}(\vec{k},\omega)$ along a momenta cut including $K$, cf. inset of Fig.~\ref{fig:MPS_dyn_AF}(a).
        }
    \label{fig:MPS_dyn_FM}
\end{figure}

\paragraph{Model B2: Honeycomb Lattice Ferromagnet.} A similar story holds for the ferromagnetic analogue of model B1. Within LSWT, the ferromagnet has two dispersive single magnon bands that meet at Dirac points at the $K$ and $K'$ points, stabilized by a spurious time-reversal symmetry \cite{mook2021, SM} originating from a U(1) in the model without DMI (group $\mathcal{G}^*_M$). As before, the DMI breaks this spurious symmetry down to $\mathcal{G}_M$ but it does not enter the linear theory; however, it leads to topological magnon gaps via magnon-magnon interactions \cite{mook2021}. Here, we confirm these topological band gaps both via real space perturbation theory and nonperturbatively using DMRG+tMPO. Using real-space perturbation theory, we derive a modified LSWT that captures the topological gap opening by Haldane-type \cite{Haldane1988} second-neighbor hoppings \cite{SM}.
The nonperturbative DMRG+tMPO data confirm that the Dirac cones are gapped out by the DMI (Fig.~\ref{fig:MPS_dyn_FM}). Moreover, they show that the topological magnon modes are not dissolved into the continuum but rather are repelled by it (in line with general arguments in Ref.~\onlinecite{Verresen2019}), pointing towards a considerably longer lifetime as expected from perturbation theory \cite{mook2021}.

\begin{figure}
    \centering
    \includegraphics[width=\linewidth,clip,trim={0 2.0cm 0 0}]{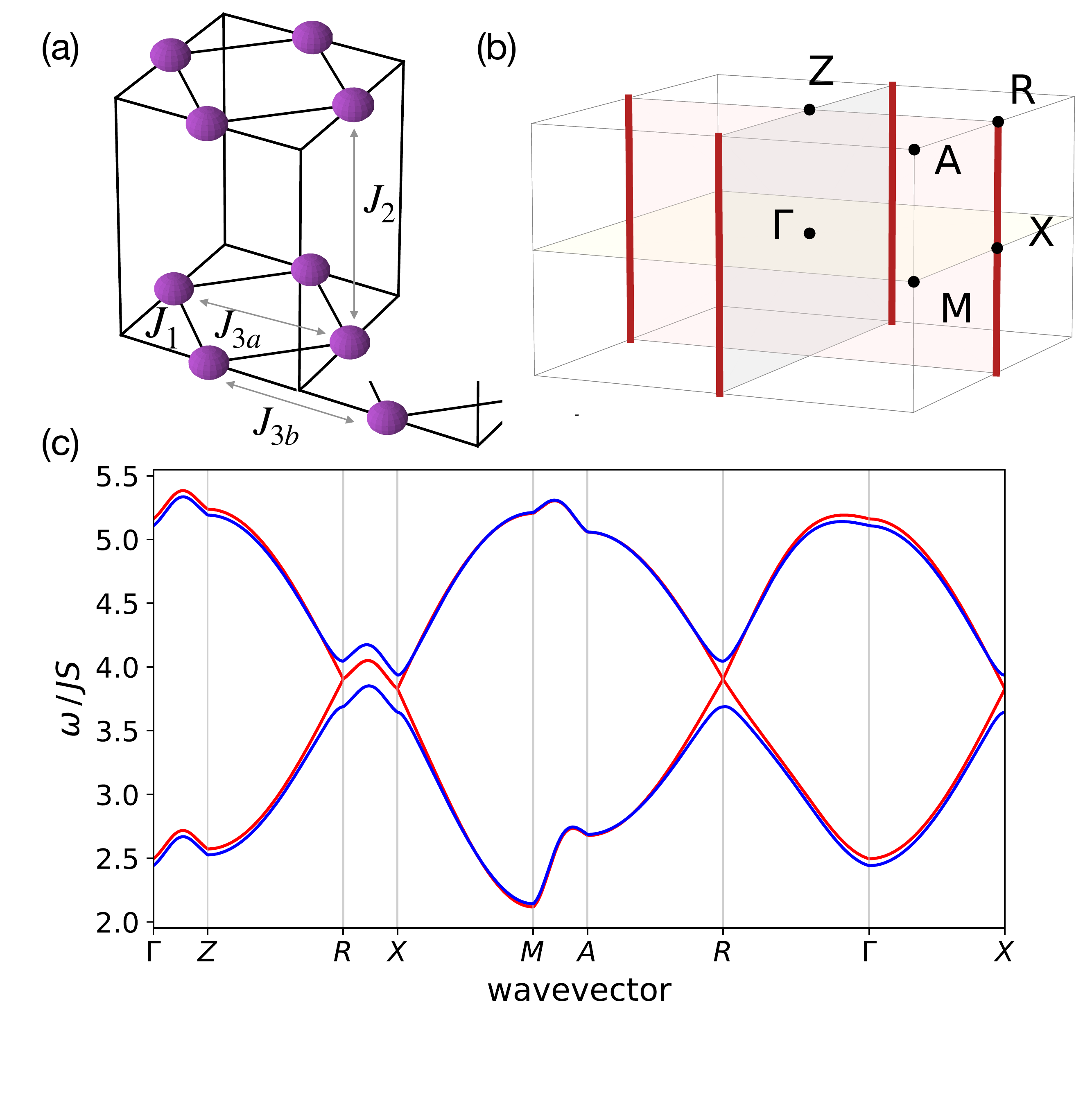}
    \caption{(a) Space group $P4$ tetragonal lattice and (b) corresponding Brillouin zone. Red lines indicate spurious nodal lines coming from the minimally coupled $J_1 + J_2$ model in LSWT.
    (c) Magnon dispersions above the ferromagnetic ground state along high symmetry directions calculated within LSWT. The $J_1 + J_2$ model (red) features spurious nodal lines, which are lifted switching on the next shell interaction $J_3$ (blue). The parameters are listed in the SM \cite{SM}.}
    \label{fig:Tetragonal_model}
\end{figure}

\paragraph{Model C: Tetragonal Lattice Model.}
The previous examples are designed to understand in a simple way the mechanisms behind spurious LSWT symmetries.
Here we show instead a non-fine-tuned 3D case, where the anisotropy blindness and shell anomaly join forces
leading to enhanced symmetry in the LSWT (captured by magnetic group $\mathcal{G}_E^*$ of Fig.~\ref{fig:squareexch}). As before, by going to further shells of interactions, the symmetry is lowered to that of the magnetic structure.
We consider a ferromagnetic bipartite lattice with a tetragonal structure described by space group $P4$ [Fig.~\ref{fig:Tetragonal_model}(a)]. This has low symmetry---with only a $C_4$ rotation around the $c$ (vertical) axis.
For complete generality, we allow all the symmetry-allowed exchange terms within a shell. 
The minimal model that connects the moments in three dimensions includes both nearest and next-nearest neighbor couplings, with a total of $14$ possible exchange terms, or $10$ if we consider the anisotropy blindness (see also \cite{SM,corticelli2022b}). We expect in this model, via representation theory, a Chern gap \cite{corticelli2022b} which nevertheless is not present in LSWT in the $J_1 + J_2$ model due to an enforced degenerate nodal line on the boundary of the Brillouin Zone [Fig.~\ref{fig:Tetragonal_model}(b)] as shown in Fig.~\ref{fig:Tetragonal_model}(c).
The expected gap is recovered by going to the next shell---the $J_1 + J_2 + J_3$ model.
The presence of this extra degeneracy can be understood in the context of spin-space group representation theory \cite{SM,corticelli2022}. The key spurious symmetry, present in the LSWT of $J_1 + J_2$, is a glide plane, which is responsible for the nodal line degeneracy. This symmetry is allowed by anisotropy blindness which, in symmetry terms, amounts to a $C_2$ spin rotation around the $c$ axis. This two-fold spin rotation remains trivial for the $J_1 + J_2 + J_3$ model. But in the $J_1 + J_2$ it combines with the remaining symmetries leading to an enhancement of the symmetries to a spin-space group.

\paragraph{Discussion and symmetry context.} Consider the hypothetical situation where one wishes to characterise the magnetism of a material from spin wave data. As will be clear from the foregoing, the implementation of symmetries in LSWT contains potential pitfalls. We offer the following practical guide to using LSWT so that no spurious degeneracies arise. 

Given a magnetic structure one may enumerate the spin and space locked transformations that leave the structure invariant. These symmetry elements form a magnetic space group $\mathcal{M}$. 
However, approximations to the full exchange Hamiltonian have the potential to break this locking leading to an enhanced symmetry formally described by spin-space groups \cite{BrinkmanElliott1966, Litvin1974}.
There may be physically well-motivated cases where 
the exchange couplings have spin-space symmetry---for example in collinear Heisenberg systems, in Kitaev magnets or generally when spin-orbit coupling is weak and there is a selection in the hierarchy of exchange terms \cite{corticelli2022}. There may be, in addition, cases where materials themselves realize a shell anomaly as a result of short-ranged couplings, leading to a physically relevant enhanced symmetry. 
This mechanism could be at work in MnF$_2$ (and possibly isostructural materials with weak spin-orbit coupling) \cite{SM,Smejkal2021altermagnets,Smejal2022,Nikotin_1969,FeF2Magnon}. Further high resolution experimental work may be of interest to look for degeneracy breaking in MnF$_2$. 


However, as we have described, there are ways in which the intended symmetries may not be represented faithfully in the excitation spectrum. For example, one may underestimate the range of significant exchange couplings in the material leading to spurious symmetries at the Hamiltonian level. In the case of anisotropy blindness, Holstein-Primakoff LSWT itself has a spurious two-fold spin rotation symmetry around the magnetization vector for a collinear system that can lead to spin-space symmetries that are absent in the exact theory. 


In general, as a practical rule of thumb, one should be especially cautious about LSWT for (i) collinear systems and for (ii) systems where the magnetic lattice has much higher symmetry than the entire crystal (model A, C), and, additionally, in the weak spin-orbit coupling regime when there are important couplings with longitudinal-transverse components (model B) \cite{SM}.
To realize the correct symmetries in LSWT one should include all couplings consistent with the fundamental symmetries out to the $n$th shell until the spectrum ceases to change qualitatively. 
As an outlook, we emphasize that, where our discussion of the shell anomaly has focussed on its realization in spin waves, the ingredients to find it may arise in tight-binding models regardless of the quasiparticle type.


\begin{acknowledgments}
\textit{Acknowledgments.}
This work was funded in part by the Deutsche Forschungsgemeinschaft (DFG, German
Research Foundation) - Project No.~504261060 (Emmy Noether Programme), SFB 1143 (project-id 247310070) and cluster of excellence ct.qmat (EXC 2147, project-id 390858490).
M.G.~acknowledges support by JSPS KAKENHI Grant Number 22K14008, 
by the Theory of Quantum Matter Unit of the Okinawa Institute of Science and Technology Graduate University (OIST),
and by the Scientific Computing section of the Research Support Division at OIST for providing the HPC resources.

\end{acknowledgments}

\bibliography{references}

\end{document}


\title{Supplementary Material for \\ ``Spurious symmetry enhancement and interaction-induced topology in magnons"}

\author{Matthias Gohlke}

\affiliation{Theory of Quantum Matter Unit, Okinawa Institute of Science and Technology Graduate University, Onna-son, Okinawa 904-0495, Japan}

\author{Alberto Corticelli}
\affiliation{Max Planck Institute for the Physics of Complex Systems, N\"{o}thnitzer Str. 38, 01187 Dresden, Germany}

\author{Roderich Moessner}
\affiliation{Max Planck Institute for the Physics of Complex Systems, N\"{o}thnitzer Str. 38, 01187 Dresden, Germany}

\author{Paul~A.~McClarty}
\affiliation{Max Planck Institute for the Physics of Complex Systems, N\"{o}thnitzer Str. 38, 01187 Dresden, Germany}

\author{Alexander Mook}

\affiliation{Institute of Physics, Johannes Gutenberg University Mainz, 55128 Mainz, Germany}

\begin{abstract}
This supplementary section contains further details on: S.I. the shell anomaly in different instances; S.II. the honeycomb lattice antiferromagnet with DMI, in particular a qualitative discussion of the effect of magnon interactions and a calculation of the spectrum from the effective model obtained using real-space perturbation theory; S.III. the honeycomb lattice ferromagnet with DMI, in particular a real-space perturbation theory to motivate second-neighbor couplings that introduce topological magnon gaps and a quantitative comparison of the induced gap as extracted from nonlinear spin-wave theory and DMRG$+$tMPO; S.IV. DMRG and time evolution numerics; S.V. a symmetry analysis of the tetragonal model put in the general context of spin-space group representation theory.
\end{abstract}

\maketitle

\section{More Details on the Shell Anomaly}

\subsection{Altermagnetic shell anomaly and M\titlelowercase{n}F$_2$}

Here we describe the shell anomaly in a physical setting: that of collinear Heisenberg antiferromagnetism in rutile structures as in MnF$_2$ and FeF$_2$.

\begin{figure}
    \centering
    \includegraphics[width=0.66\linewidth, clip, trim = 0 50 0 50]{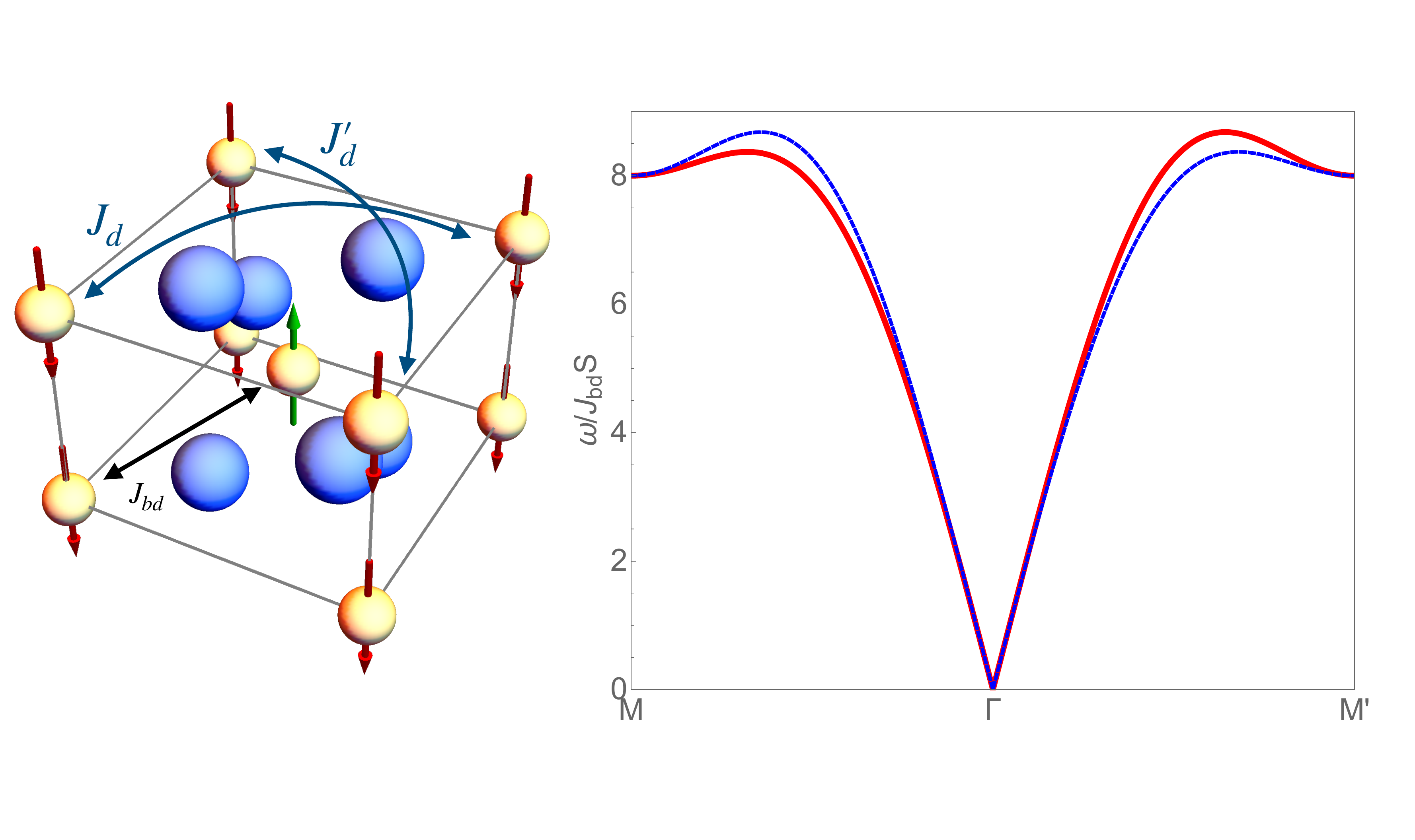}
    \caption{Left panel: crystal structure of MnF$_2$. Yellow ions are Mn(II) in a body centered tetragonal arrangement and fluoride ions are in blue. The structure has a $C_{2z}$ and a $C_{2}$ about $[110]$ and $[1\bar{1}0]$. In addition there are nonsymmorphic (i) $C_{4z}$ combined with a translation through half of the body diagonal and (ii) $C_{2}$ about $[010]$ with the same translation. Right panel: spectrum along a high symmetry direction with $J_{bd}=1$, $J_d=-0.2$ and $J'_d=-0.3$ [for the linear spin wave model of Eq.~\eqref{eq:mnf2}]. The colors of the modes indicate the d-wave nature of the splitting.}
    \label{fig:MnF2}
\end{figure}

MnF$_2$ has a tetragonal crystal structure of fairly low symmetry (space group 136, P4$_2$/mnm) with a body-centred arrangement of magnetic manganese ions  and fluoride ions in a crisscross pattern [see Fig.~\ref{fig:MnF2}(left)]. Below the N\'{e}el temperature of {$66.5$~K} the structure is a simple collinear antiferromagnetic structure with a weak easy axis anisotropy along the $c$ direction \cite{MnF2}. Fits to triple axis inelastic neutron scattering \cite{Nikotin_1969} data give the corner to center exchange $J_{bd}$ as the dominant magnetic coupling followed by the nearest neighbor coupling along the $c$ direction. Single ion terms were deemed negligible and the anisotropy direction was argued to originate from the comparatively weak dipolar coupling. The experiment did not resolve a splitting between the magnon bands along the measured directions: (h00), (00l), (h0h), (1/2,0,l), (h,0,1/2) \cite{Nikotin_1969}. There is a similar apparent absence of splitting in FeF$_2$ \cite{FeF2Magnon}.

Here we point out that the double degeneracy of the magnon bands is lifted by isotropic exchange despite the body-centered tetragonal structure essentially because of the low symmetry of the structure. For example, the exchange couplings (labelled $J_d$, $J'_d$ in the figure) across in-plane diagonals are not constrained to be identical along $[110]$ and $[1\bar{1}0]$. The model is
\begin{align}
H_{\vec{k}} & = \left( \begin{array}{cccc} 
A_{\vec{k}} & 0 & 0 & B_{\vec{k}} \\
0 & \bar{A}_{\vec{k}} & B_{\vec{k}} & 0 \\
0 & B_{\vec{k}} & A_{\vec{k}} & 0 \\
B_{\vec{k}} & 0 & 0 & \bar{A}_{\vec{k}} 
\end{array} \right)  \label{eq:mnf2} \\
A_{\vec{k}} & = 8J_{\rm bd} - 2(J_d + J'_d) + 2J_d\cos(k_x - k_y) + 2J'_d\cos(k_x + k_y) \nonumber \\
\bar{A}_{\vec{k}} & = 8J_{\rm bd} - 2(J_d + J'_d) + 2J'_d\cos(k_x - k_y) + 2J_d\cos(k_x + k_y) \nonumber \\
B_{\vec{k}} & = J_{bd}\sum_{i \in bd} \nonumber \exp(\mathrm{i}\vec{k}\cdot \vec{\delta}_{i})
\end{align}
where $\vec{\delta}_{i}$ are the eight body-diagonals $1/2(1,1,1)$ etc.

In the right panel of Fig.~\ref{fig:MnF2} we show the effect of the symmetry-allowed Heisenberg exchange showing that the degeneracy is lifted along directions of low symmetry. In order to detect this splitting it is not sufficient to explore the directions measured in the 1969 experiment \cite{Nikotin_1969}. If splitting of this nature is present in the material it should be most visible along the $(hh0)$ direction. It may be interesting to follow this up with new experiments on a high resolution instrument. If splitting is resolved the analysis would require a detailed study of possible weak anisotropies in the magnetic Hamiltonian in order to establish the origin of any observed splitting. This phenomenon of degeneracy breaking in simple antiferromagnets with Heisenberg exchange is identical to so-called altermagnetism \cite{Smejkal2021altermagnets,Smejal2022} where spin splitting occurs in electronic band structures despite negligible spin-orbit coupling as a result of low lattice symmetry. In this case, this is reflected in a d-wave ``spin'' splitting meaning that the nature of the magnon wavefunctions is swapped in energy in going from $(h,h,0)$ to $(h,-h,0)$. 

\subsection{Further remarks on the shell anomaly}

In the previous section and in the main text we discussed an example where the shell anomaly takes the form of a degeneracy at the nearest neighbor level that is lifted or broken by further neighbor couplings and that the nature of the degeneracy breaking is into a d-wave ``spin''-split altermagnet. The resolution of a shell anomaly can, however, take several forms.

\begin{figure}
    \centering
    \includegraphics[width=0.3\linewidth,trim={0 0 0 0},clip]{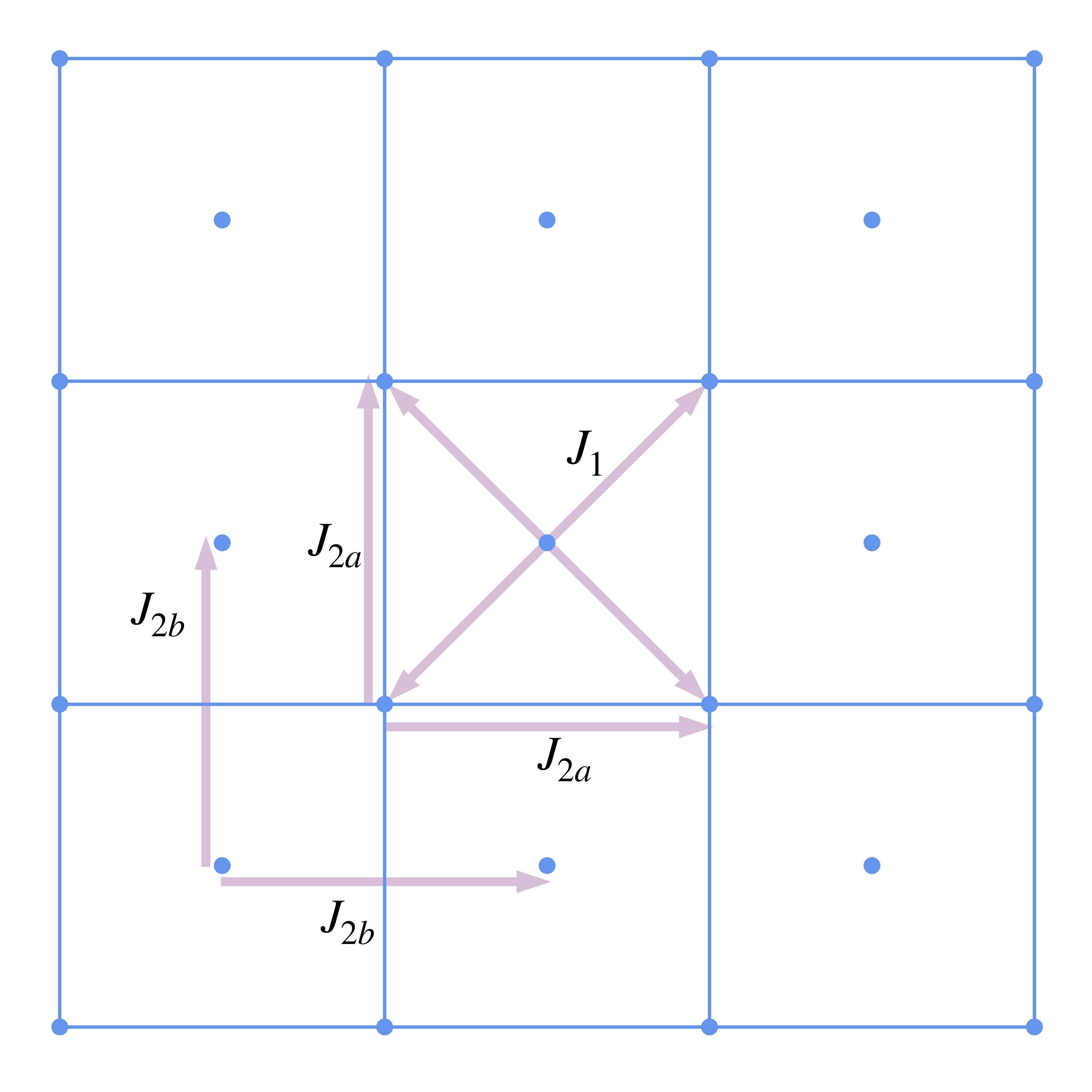}
    \caption{Square lattice with various exchange paths indicated used to illustrate a simple case where spin wave theory can fail to capture the symmetries of the problem.}
    \label{fig:squareexch}
\end{figure}

Here is another example. Figure~\ref{fig:squareexch} shows a two sublattice structure. We suppose that the lattice has a $C_4$ symmetry about the square centres and that the exchange is rotationally symmetric. The symmetries ensure that there is a single nearest neighbor coupling $J_1$ on all the bonds joining sublattices $a$ and $b$ in a primitive cell. If this is antiferromagnetic, the model is identical to the simple square lattice antiferromagnet with a double degeneracy in the magnon spectrum. However, the lattice symmetries do not include a sublattice symmetry between $a$ and $b$ and further neighbor exchange will, in general, lift the double degeneracy. Specifically, if the $J_2$ couplings are different on the two sublattices the sublattice symmetry present in the nearest neighbor model [cf.~Fig.~\ref{fig:squareexch}] is lifted. As in the altermagnetic case, the exact spin wave theory has an enhanced symmetry at the nearest neighbor level that is lifted by further neighbor couplings. In fact the magnon spectrum is not the only, or even the main, effect of the further neighbor couplings. Because the two sublattices are inequivalent in this model, fluctuations will reduce the moments on the sublattices unequally leading to a net magnetization. So the lifting of the shell anomaly in this case takes an antiferromagnet into a ferrimagnet.

\section{Honeycomb-Lattice Antiferromagnet}
\subsection{Qualitative discussion of magnon-magnon interactions}

Within spin-wave theory, one performs a Holstein-Primakoff transformation from spins to bosons, which reads
\begin{align}
    \hat{\vec{S}}_{i} 
    = 
    \sqrt{\frac{1}{2}} 
    \left( 
    	\sqrt{ 1- a_i^\dagger a_i } a_i \vec{e}^{-} 
    	+ 
    	a_i^\dagger \sqrt{ 1- a_i^\dagger a_i } \vec{e}^{+} \right) 
    + \left( \frac{1}{2} - a_{\vec{r}}^\dagger a_{\vec{r}}\right) \vec{e}^{z},
    \label{eq:HPexpansion}
\end{align}
for $S=1/2$. The bosonic operators obey the usual commutation relation 
$
	[a_i,a^\dagger_j] = \delta_{i,j} 
$. 
The axes of the local reference frame span an orthonormal basis $\{\vec{e}^x, \vec{e}^y, \vec{e}^z \}$, with $\vec{e}^\pm = (\vec{e}^x \pm \mathrm{i} \vec{e}^y)/\sqrt{2}$ and $\vec{e}^z$ along the classical ground state direction.

Upon plugging Eq.~\eqref{eq:HPexpansion} into the spin Hamiltonian of the Heisenberg antiferromagnet (see main text), an expansion about the classical N\'{e}el order yields
\begin{align}
    H = H_0(J_z) + H_2(J,J_z) + H_3(D) + H_4(J,J_z) + \ldots,
\end{align}
where $H_0(J_z)$ is an irrelevant constant. Within \emph{linear} spin-wave theory, we diagonalize the quadratic piece $H_2(J,J_z)$ to obtain the magnon normal modes. Interactions between magnons are found in the three-magnon term $H_3(D)$, the four-magnon term $H_4(J,J_z)$, and higher terms.

Importantly, DMI does not enter the harmonic theory and instead appears, to lowest order, in the cubic Hamiltonian $H_3(D)$. As was shown in the main text, this leads to a spurious magnon band degeneracy over the entire Brillouin zone in the linear theory and the magnons carry a spurious good spin quantum number $\sigma = \uparrow, \downarrow$. The linear spin-wave theory is blind to the spin-nonconserving effect of the DMI.

\begin{figure}
    \centering
    \includegraphics[width = 1\columnwidth]{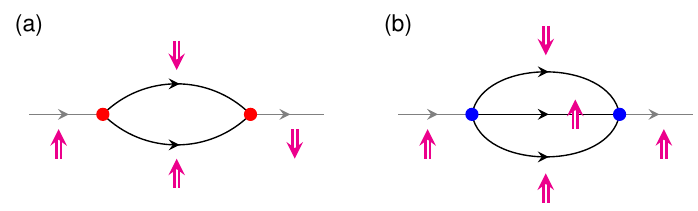}
    \caption{Sketches of selected zero-temperature self-energy Feynman diagrams as encountered in nonlinear spin-wave theory. Thin black arrows indicate propagators and magenta arrows the spin of magnons. The colored circles denote interaction vertices; red circles are DMI-induced three-magnon vertices and blue circles are exchange-induced four-magnon vertices. (a) DMI-induced bubble diagram that facilitates spin flips. Note that in-coming and out-going magnons have opposite spin. (b) Exchange-induced bubble diagram which conserves the magnon spin because the XXZ-type exchange term comes with a $U(1)$ symmetry.}
    \label{fig:diagram}
\end{figure}

In the main text, we demonstrate how a real-space perturbation theory captures the spin-nonconserving effects of DMI and the resulting splitting of the magnon bands. Here, we complement this analysis by a qualitative discussion how the spin-splitting occurs in a nonlinear spin-wave theory. By treating $H_3(D)$ and $H_4(J,J_z)$ as perturbations to $H_2(J,J_z)$, a many-body perturbation theory can be carried out. Figure \ref{fig:diagram} shows two examples of Feynman diagrams that contribute to the magnon self-energy at zero temperature. To second-order in $H_3(D)$, which yields a $1/S$ correction to the bare magnon energies, one encounters the diagram in Fig.~\ref{fig:diagram}(a). The three-magnon vertices derive from the spin-nonconserving nature of the DMI and, hence, allow for the total spin to change, giving rise to a spin-flip self-energy that mixes spin-$\uparrow$ and spin-$\downarrow$ magnons. Consequently, the poles of the renormalized magnon Green's function are no longer degenerate and the renormalized magnon energies split. This finding is to be contrasted with a second-order perturbation in $H_4(J,J_z)$, which is a $1/S^2$ correction to the bare magnon energies. Without DMI, the spin Hamiltonian $H$ is $U(1)$ symmetric. Thus, although the magnon particle number is not conserved---allowing for one-in-three-out vertices (and \textit{vice versa}) that make up the diagram in Fig.~\ref{fig:diagram}(b)---the magnon spin is conserved \cite{Harris1971}.
As a result, the self-energy does not mix magnon spin species and the magnon band degeneracy stays intact.

\subsection{Effective model from real-space perturbation theory}
We write the spin Hamiltonian in Eq.~(1) of the main text as $H = H_0 + V$, with the unperturbed piece
\begin{align}
    H_0 = \frac{J_z}{2} \sum_{\langle i j \rangle} S_i^z S_j^z
\end{align}
and the perturbation $V = V_D + V_J$, where
\begin{align}
    V_J = \frac{J}{4} \sum_{\langle i j \rangle} \left( S_i^+ S_j^- + S_i^- S_j^+ \right)
\end{align}
and
\begin{align}
    V_D = \frac{1}{2} \sum_{\langle i j \rangle} \vec{D}_{ij} \cdot \vec{S}_i \times \vec{S}_{j}.
\end{align}
For $J_z>0$, the antiferromagnetic N\'{e}el order is the exact ground state of the Ising model $H_0$. We work in the local reference frame and replace $S_i = R_i \tilde{S}_i$, where $R_i$ is the identity for the spin-up (A) sublattice and $R_i = {\rm diag}(1,-1,-1)$ for the spin-down (B) sublattice, leading to $S^z_i = - \tilde{S}^z_i$ and $S^\pm_i = \tilde{S}^{\mp}_i$. In the local frame, $V_J$ features double flips, $\tilde{S}_i^-\tilde{S}_j^-$, while $V_D$ contains transverse-longitudinal couplings, $\tilde{S}_i^{\pm}\tilde{S}_j^z$.

We denote the fully sublattice polarized N\'{e}el state by $|\tilde{0}\rangle$ and consider single-flip states $|\tilde{i} \rangle = \tilde{S}_i^- |\tilde{0}\rangle$, with energy
$
    E_i 
    =
    \langle \tilde{i} | H_0 | \tilde{i} \rangle
    =
    3 J_z / 2.
$
To first order in the perturbation, we find no coupling because $V$ in the local frame does not conserve the number of spin flips. Instead, hopping is a second-order process. First, we discuss $V_J$. 
The pair creation and destruction process via states with three spin-flips (energy $E_v = 5 J_z/2$) generates an effective intra-sublattice second-neighbor hopping,
\begin{align}
 	\frac{-1}{J_z} 
 	\left\langle \singlek \right| 
 	V_J
 	\left| \triple \right\rangle 
 	\left\langle \triple \right| 
 	V_J 
 	\left| \singlei \right\rangle 
 	\propto
     \frac{-J^2}{J_z},
\end{align}
where $i$ and $k$ belong to the same sublattice and the states depict the pattern of spin flips generated by $V_J$. White (black) circles indicate the ground state (spin flips).
In the global frame, such a hopping would read $\propto \frac{J^2}{J_z} S^+_i S^-_k$ in the spin Hamiltonian. Thus, as far as particle hopping is concerned, the honeycomb lattice decomposes into two interwoven but noninteracting triangular-lattice ferromagnets with opposite magnetization direction. This decomposition ensures spin conservation and the sublattice symmetry results in a doubly degenerate spectrum. This result is qualitatively similar to just carrying out a linear spin-wave theory of the full spin Hamiltonian in Eq.~(1) of the main text.

As we have emphasized in the main text, a hopping between sublattices is generated to second order in $V_D$:
\begin{align}
	\frac{-2}{J_z} \left\langle \singlej \right| 
	V_D 
	\left| \doubleij \right\rangle 
	\left\langle \doubleij \right| 
	V_D 
	\left| \singlei \right\rangle 
	& \propto
	\frac{D^2}{J_z} \mathrm{e}^{-2\mathrm{i} \varphi_{ij}},
	\nonumber
\end{align} 
where $\tan \varphi_{ij} = D_{ij}^y/D_{ij}^x$.
In the global frame, this coupling mimicks bond-dependent off-diagonal exchange interaction. Thus, in the perturbative limit, we replace DMI by bond-dependent off-diagonal exchange, resulting in Eq.~(2) of the main text, which is reproduced here for convenience:
\begin{align}
    H'
    = 
    \sum_{i \in \text{A}} \sum_{j = 0}^2 &\left[ 
    J_z S_i^z S_{i+\vec{\delta}_j}^z + \frac{J}{2} \left( S_i^+ S_{i+\vec{\delta}_j}^- + S_i^- S_{i+\vec{\delta}_j}^+ \right)
    + \frac{J_{++}'}{2} \left( \mathrm{e}^{\mathrm{i} \vartheta_{\vec{\delta}_j}} S_i^+ S_{i+\vec{\delta}_j}^+ + \mathrm{e}^{-\mathrm{i} \vartheta_{\vec{\delta}_j}} S_i^- S_{i+\vec{\delta}_j}^- \right)  
    \right], 
    \quad \vartheta_{\vec{\delta}_n} = 2\pi n /3\label{eq:HamiltonianAFM}
\end{align}

\begin{figure}
    \centering
    \includegraphics[width = 0.4\columnwidth]{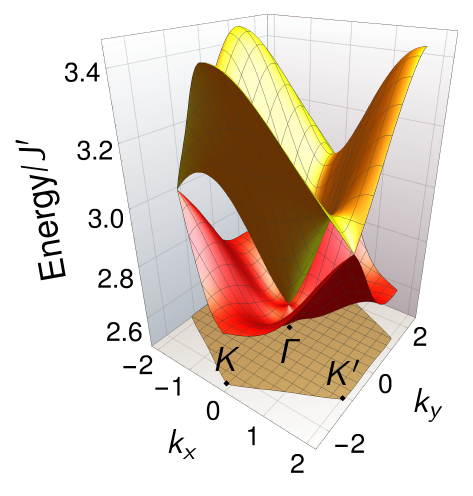}
    \caption{Sketch of the magnon band structure (red and yellow bands) as obtained from diagonalizing $H'_{\vec{k}}$ in Eq.~\eqref{eq:effectiveH}, with $J=1$, $J_z = 2$, and $J'_{\pm\pm} = 0.3$. The lower brown hexagon indicates the Brillouin zone. Clearly, the magnon bands are split except for the $\Gamma$ and the $K'$ points that support Dirac magnons.}
    \label{fig:effectiveBS}
\end{figure}

Below, we provide the details of the LSWT of spin Hamiltonian $H'$ in Eq.~\eqref{eq:HamiltonianAFM}.
After expanding in bosons, a bilinear Hamiltonian
$
    H'_2(J,J_z,D) 
    = 
    \frac{1}{2} \sum_{\vec{k}} \vec{\psi}^\dagger_{\vec{k}} H'_{\vec{k}} \vec{\psi}_{\vec{k}}
$
is obtained that no longer features a block-diagonal kernel. Instead, the kernel reads
\begin{align}
    H'_{\vec{k}} 
    = 
    \frac{1}{2} \begin{pmatrix}
    	3 J_z & - J \gamma_{\vec{k}}  & - J_{++}' \lambda_{\vec{k}} & 0 \\
    	- J \gamma_{-\vec{k}} & 3 J_z & 0 & - J_{++}' \lambda_{-\vec{k}}  \\
    	 - J_{++}' \lambda^\ast_{\vec{k}} & 0 & 3 J_z & - J \gamma_{-\vec{k}} \\
    	0 &  - J_{++}' \lambda^\ast_{-\vec{k}} & - J \gamma_{\vec{k}} & 3 J_z \\
    \end{pmatrix},
    \quad
    \lambda_{\vec{k}}
	=
	\sum_{i=0}^2 \mathrm{e}^{\mathrm{i} \left( \vec{k} \cdot \vec{\delta}_i -\vartheta_{\vec{\delta}_i} \right)},
	\label{eq:effectiveH}
\end{align}
where $\gamma_{\vec{k}}$ and $\vec{\delta}_i$ are given in the main text.
After paraunitary diagonalization of $H'_{\vec{k}}$ the magnon spectrum is obtained; a representative example is depicted in Fig.~\ref{fig:effectiveBS}. The degeneracy of the magnon modes is lifted throughout the Brillouin zone except for the $\Gamma$ and the $K'$ point, where $\lambda_{\vec{k}}$ is zero, resulting in Dirac cones. The magnon band splitting is maximal at the $K$ point where it reaches $3|J'_{++}|$. We note that flipping the N\'{e}el vector flips the role of the $K$ and $K'$ points.

\section{Honeycomb-Lattice Ferromagnet: Nonlinear spin-wave theory}
In the main text, we contrasted the antiferromagnetic case with the ferromagnetic case by providing a short summary of the most important features. Here, we present the details.
We consider the honeycomb-lattice spin-$1/2$ Hamiltonian
\begin{align}
	H 
	= 
	\frac{1}{2} \sum_{\langle i j \rangle} \left[ -J \vec{S}_i \cdot \vec{S}_j + \vec{D}_{ij} \cdot \left( \vec{S}_i \times \vec{S}_j \right) \right] - B \sum_i S_i^z,
\end{align}
with $J>0$ being nearest-neighbor Heisenberg exchange interaction, $\vec{D}_{ij} = D \hat{\vec{z}} \times \hat{\vec{e}}_{ij}$ being interfacial DMI; $\hat{\vec{e}}_{ij}$ is a unit vector along bond direction and $\hat{\vec{z}}$ along the lattice normal. For large enough external field $B$, spiralization is overcome and the ground state is the fully polarized state.
Within spin-wave theory, we perform the Holstein-Primakoff transformation in Eq.~\eqref{eq:HPexpansion}.
The axes of the local reference frame read $\vec{e}^\pm = ( 1, \pm\mathrm{i}, 0 )/\sqrt{2}$, and $\vec{e}^{z} = (0,0,1)$.
After expanding the square roots, we obtain
\begin{align}
    H = H_0(J,B) + H_2(J,B) + H_3(D) + H_4(J) + \ldots
\end{align}
Importantly, DMI enters again to lowest order in the cubic Hamiltonian $H_3(D)$, rendering the harmonic theory blind to DMI. Indeed, the harmonic Hamiltonian reads
$
    H_2(J,B) 
    = 
    \sum_{\vec{k}} \vec{\psi}^\dagger_{\vec{k}} H_{\vec{k}} \vec{\psi}_{\vec{k}}
$, 
with 
$
    \vec{\psi}_{\vec{k}}^\dagger = (a^\dagger_{\vec{k},1}, a^\dagger_{\vec{k},2})
$ 
being the vector of Fourier transformed sublattice magnon creators and
\begin{align}
    H_{\vec{k}} = 
    \frac{1}{2}
    \begin{pmatrix}
		3 J + B & -J\gamma_{\vec{k}} \\
		-J\gamma_{-\vec{k}} & 3J + B
	\end{pmatrix},
	\quad
	\gamma_{\vec{k}} = \sum_{i=0}^2 \mathrm{e}^{\mathrm{i} \vec{k} \cdot \vec{\delta}_i},
	\label{eq:HamiltonMarix}
\end{align}
the spin-wave matrix.
The nearest-neighbor bonds $\vec{\delta}_i$ are identical to those defined in the main text for the antiferromagnetic model.
After a diagonalization, we find the energies of the normal mode magnons
\begin{align}
	\varepsilon_{\vec{k},\pm} 
	= 
	\frac{J}{2}\left( 3 \pm |\gamma_{\vec{k}}| \right) + B. 
	\label{eq:single-magnon}
\end{align}
There are two branches, which are degenerate at the Brillouin zone corners $\pm\vec{K}$ where $|\gamma_{\pm\vec{K}}| = 0$. The resulting Dirac cones are spurious because linear spin-wave theory does not account for the time-reversal symmetry breaking effects of DMI \cite{mook2021}.

The cubic Hamiltonian reads
\begin{align}
    H_3(D) = 
    \frac{1}{2 \sqrt{N}} \sum_{l,m,n = 1}^2 \sum_{\vec{k}, \vec{q}, \vec{p}}^{\vec{p}=\vec{k}+\vec{q}} \left(
         V^{lm \leftarrow n}_{\vec{k}, \vec{q} \leftarrow \vec{p}}(D)
        a^\dagger_{\vec{k}, l} a^\dagger_{\vec{q}, m}  a_{\vec{p}, n} 
        +
        \mathrm{H.c.} \right),
\end{align}
where we made the DMI dependence of the three-magnon vertex explicit. To account for the effects of cubic interactions on the single-particle excitations, we evaluate the self-energy $\Sigma(\omega,\vec{k})$ of the single-bubble Feynman diagram (for technical details, see Ref.~\onlinecite{mook2021}) and extract the poles of the renormalized single-particle Green's function $G(\omega,\vec{k}) = [\omega - H_{\vec{k}} - \Sigma(\omega,\vec{k})]^{-1}$. It can be shown that $\Sigma(\omega,\vec{k})$ generates a mass term $\propto \sigma_3$ (in the sublattice basis) that gaps out the Dirac magnons; this gap was shown to be topologically nontrivial by deriving an effective Hamiltonian in Ref.~\onlinecite{mook2021}. 

Below, we go beyond Ref.~\onlinecite{mook2021} by comparing the results of nonlinear spin-wave theory to that of (1) the nonperturbative DMRG+tMPO and (2) a real-space perturbation theory. These methods have different merits: DMRG+tMPO proves that a nonperturbative treatment confirms the predictions of perturbation theory and the real-space perturbation theory explains why the interaction-induced magnon gap is topologically nontrivial.

\subsection{Small magnetic fields ($B/J \sim 1$)}
According to Eq.~\eqref{eq:single-magnon}, the single-magnon energies at the $\pm \vec{K}$ points are given by $\varepsilon_{\vec{K},\pm} = 3J/2 + B$. Since the DMI-induced three-magnon interactions facilitate an interaction with two-magnon states, $\varepsilon_{\vec{K},\pm}$ has to be compared with the lower threshold of the two-magnon continuum, $\varepsilon_{\vec{K}}^{(2)} = J + 2B$, to judge whether the single-particle modes fall inside or outside the continuum. There are two cases:
\begin{enumerate}
    \item $\varepsilon_{\vec{K},\pm} > \varepsilon_{\vec{K}}^{(2)} \; \iff  \; B < J/2$: The single-magnon modes fall inside the continuum and are kinematically allowed to decay into two other magnon modes that make up the continuum.
    \item $\varepsilon_{\vec{K},\pm} < \varepsilon_{\vec{K}}^{(2)} \; \iff  \; B > J/2$: The single-magnon modes fall outside the continuum and are kinematically forbidden to decay. Hence, the renormalization of the single-magnon modes is purely real.
\end{enumerate}
Above, we determined the threshold of the two-magnon continuum by looking at the sum of two \emph{bare} magnon energies. Self-consistency would require that, as the single-magnon energies get renormalized, so does the threshold of the continuum. Indeed, nonperturbative calculations show that, in general, the two-magnon continuum states repel the single-particle energies, such that lifetime broadening is considerably suppressed \cite{Verresen2019}. Thus, the above distinction between the two cases is approximate and only applies to lowest-order perturbation theory. Still, the estimate $B = J/2$ provides a good rule of thumb for where to expect the largest renormalization effects.

\begin{figure}
    \centering
    \includegraphics[width=0.41\columnwidth]{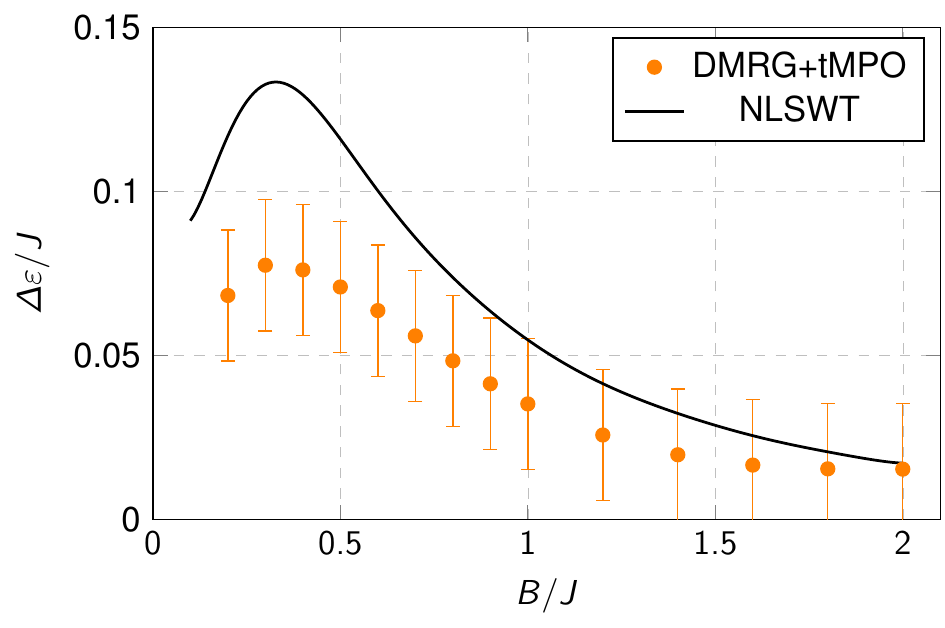}
    \caption{
    Interaction-induced Dirac magnon gap $\Delta \varepsilon$ in the spin-$1/2$ honeycomb ferromagnet as a function of magnetic field $B$, as extracted from nonlinear spin-wave theory (NLSWT, black line) and DMRG+tMPO (orange points). 
    Error bars indicate the systematic width of the Gaussian peak in the dynamical structure factor (see Sec.~\ref{sec:technicalitiesFM} for technical details).
    }
    \label{fig:DMRG}
\end{figure}

We now turn to the comparison of nonlinear spin-wave perturbation theory to a nonperturbative treatment using DMRG+tMPO. In both cases, we concentrate on the renormalization-induced magnon band gap $\Delta \varepsilon$ at the $\pm \vec{K}$ points, that is to say, on the splitting of the spurious harmonic Dirac magnons. 
In DMRG+tMPO, we extract $\Delta \varepsilon$ from the splitting of the intensity peaks in the dynamical spin structure factor (see main text). In perturbation theory, we extract it from the poles of the renormalized magnon Green's function as extracted from an off-shell solution of the Dyson equation \cite{mook2021} to capture the repulsion between single-particle states and the continuum. 

Figure \ref{fig:DMRG} depicts the Dirac magnon gap $\Delta \varepsilon$ as a function of $B$ as obtained from the above procedure. We find excellent qualitative agreement. Coming from large fields, where the gap is suppressed because of the energetic distance between single-particle states and the two-magnon continuum, both methods capture the growing gap upon lowering $B$. The maximal splitting is found for fields just below $B = J/2$, as expected from the naive argument. Further decreasing the field leads to a suppression of the gap because the continuum pushes the two magnon branches together. Quantitatively, we find that perturbation theory overestimates the gap but still provides a good order of magnitude estimate.

\subsection{Large magnetic fields ($B/J \gg 1$)}
We now concentrate on the opposite case of large magnetic fields $B/J \gg 1$. In this limit, the finite numerical linewidth broadening in DMRG+tMPO prevented us from resolving the band gap. Thus, instead we develop a real-space perturbation theory to explain the results of the nonlinear spin-wave theory.

\begin{figure}
    \centering
    \includegraphics[scale=0.85]{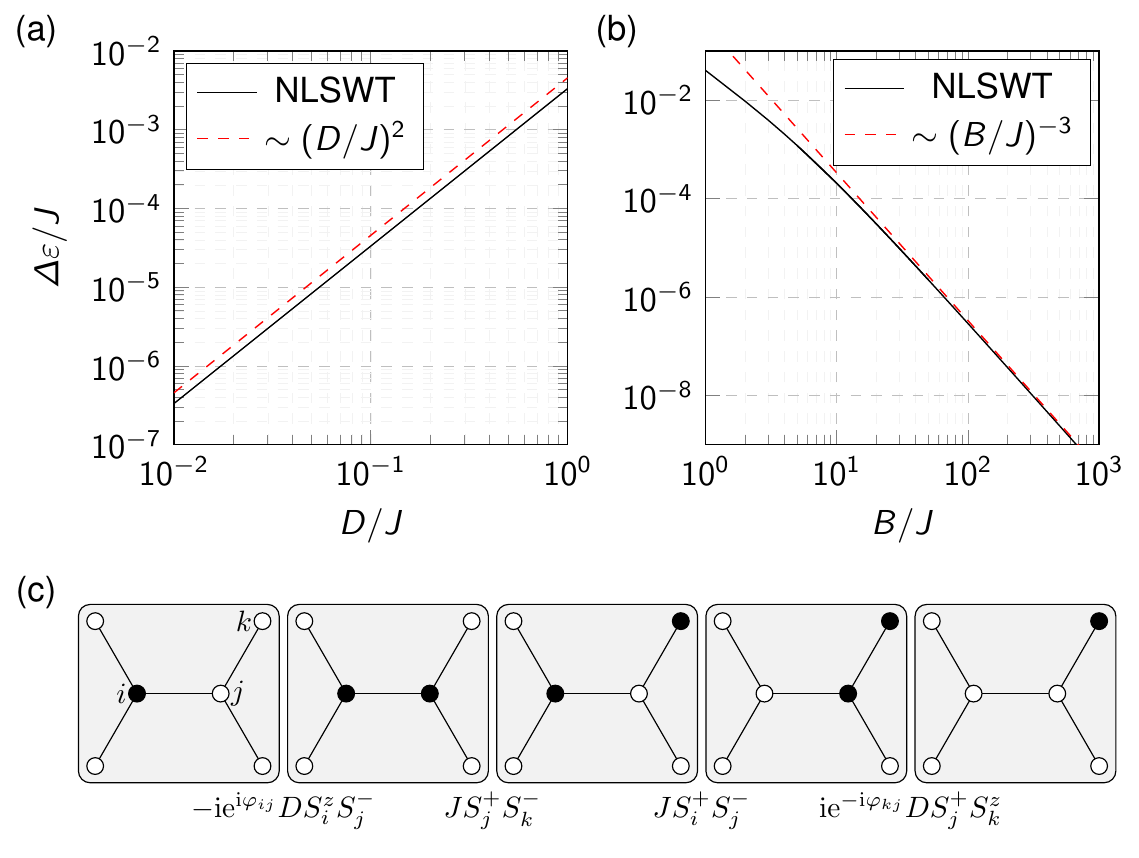}
    \caption{
    Interaction-induced Dirac magnon gap in the spin-$1/2$ honeycomb ferromagnet in the limit of large fields.
    (a,b) Magnon gap $\Delta \varepsilon$ versus DMI $D$ and magnetic field $B$, as obtained from nonlinear spin-wave theory (NLSWT) in the limit $D,J \ll B$. Parameters read (a) $B/J=5$ and (b) $D/J=0.6$.
    (c) Sketch of a real-space perturbation process generating time-reversal-symmetry-breaking second-neighbor coupling. Panels represent basis states and have to be read from left to right. White circles indicate spins in the ferromagnetic ground state, black circles indicate flipped spins. Between neighboring panels, we indicate which perturbing operators are sandwiched.  
    }
    \label{fig:FM}
\end{figure}

Figures~\ref{fig:FM}(a) and (b) show the scaling of the magnon band gap $\Delta \varepsilon$ at the $\pm \vec{K}$ points as a function of $D$ and $B$, as obtained from nonlinear spin-wave theory. As expected for a second-order process in DMI, $\Delta \varepsilon \propto D^2$. However, the magnetic-field dependence $\Delta \varepsilon \propto 1/B^3$ at large $B$ is curious because one might naively expect a $1/B$ scaling of a second-order self-energy.

For a real-space perturbation theory in the limit of large fields, we separate the spin Hamiltonian $H = H_0 + V$ into an unperturbed piece $H_0 = - B \sum_i S_i^z$ and a perturbation $V = V_J + V_D$, with
\begin{align}
    V_J &= - \frac{J}{2} \sum_{\langle ij \rangle} \vec{S}_i \cdot \vec{S}_{j} 
    ,
    \\
    V_D &= \frac{1}{2} \sum_{\langle ij \rangle}
         \vec{D}_{ij} \cdot \vec{S}_i \times \vec{S}_{j} 
         = 
         \frac{D}{4} \sum_{\langle ij \rangle}
		 \left[    
		 -\mathrm{i}\mathrm{e}^{-\mathrm{i}\varphi_{ij}}	S^+_i S^z_j	 
		 +\mathrm{i}\mathrm{e}^{\mathrm{i}\varphi_{ij}}  S^-_i S^z_j +\mathrm{i}\mathrm{e}^{-\mathrm{i}\varphi_{ij}}  S^z_i S^+_j
		 -\mathrm{i}\mathrm{e}^{\mathrm{i}\varphi_{ij}}  S^z_i S^-_j
         \right],
\end{align}
where $\tan \varphi_{ij} = D_{ij}^y/D_{ij}^x$. Explicitly, by rewriting the angles $\varphi_{ij}$ such that they are given along nearest-neighbor bonds, e.g., $\varphi_{ij} = \varphi_{\vec{\delta}_0}$ for $\vec{r}_j-\vec{r}_i = \vec{\delta}_0$, we find $\varphi_{\vec{\delta}_n} = 2\pi n/3 - \pi $.
The sub-Hilbert space of interest contains the single spin-flip states $|i\rangle = S_i^- |0\rangle$, where $|0\rangle$ is the fully polarized ferromagnetic ground state. The matrix elements of an effective Hamiltonian read \cite{CohenTannoudji1998}
\begin{align}
	H^\text{eff}_{ij}
	=	
	\langle i | H_0 | j \rangle
	+
	\langle i | V | j \rangle
	+
	\frac{1}{2} \sum_{v} \langle i | V | v \rangle \langle v | V | j \rangle \left( \frac{1}{E_i - E_v} + \frac{1}{E_j - E_v}  \right)
    + \text{higher-order terms}.
\end{align} 
Virtual states $|v\rangle = |m,n\rangle =  S^-_{m} S^-_{n} |0\rangle$ contain two spin flips ($m \ne n$). Below, we give energies relative to the ground state energy $E_0 = \langle 0 | H_0 |0 \rangle = -NB/2$, where $N$ is the total number of spins. (We do not explicitly denote the subtracted ground state energy.)

1) The unperturbed energies of the single spin-flip states are determined by $\langle i | H_0 | j \rangle = B \delta_{i,j}$. We define $E_i \equiv \langle i | H_0 | i \rangle = B$.

2) To first order in the perturbation, we obtain
\begin{align}
    \langle j | V | i \rangle 
    =
    \langle j | V_J | i \rangle 
    = 
    \frac{3J}{2} \delta_{i,j} + \frac{J}{2} \delta_{j,i+\delta},
    \label{eq:ferro-first-order}
\end{align}
where $i+\delta$ is shorthand for $\vec{r}_i + \vec{\delta}$ and $\vec{\delta}$ drawn from the three nearest-neighbor bond vectors. From Eq.~\eqref{eq:ferro-first-order}, we find that there is an on-site potential and a spin flip hopping between nearest neighbors.

3) In second order, we consider transitions via virtual two-magnon states $|v\rangle$, whose energy is $E_v \equiv \langle v | H_0 | v \rangle = 2 B$. Such a transition is only possible by a double-DMI process, i.e.,
\begin{align}
    -\frac{1}{B} \sum_{v} \langle j | V_D | v \rangle \langle v | V_D | i \rangle 
    \propto
    - \frac{D^2}{B} \delta_{i,j} 
    + \frac{D^2}{B} \delta_{j,i+\delta}.
\end{align}
It provides a correction to the on-site potential and the spin-flip hopping. The negative sign in front of $\delta_{i,j}$ indicates a level repulsion between single-magnon states and the two-magnon continuum, in agreement with the general considerations of Ref.~\onlinecite{Verresen2019}. 

4) So far, no gap opening term was encountered. Such a process involves an effective second-nearest neighbor hopping $t_2$ for the single spin-flip states, as depicted in Fig.~\ref{fig:FM}(c). This process is found only at fourth order in the perturbation,
\begin{align}
	t_2 
	\sim
	\frac{	
		\langle k | V_D |j,k\rangle
		\langle j,k | V_J |i,k\rangle
		\langle i,k | V_J |i,j\rangle
		\langle i,j | V_D |i \rangle
	}
	{
		(E_i - E_v)^3	
	},
\end{align}
establishing its scaling with $1/B^3$. Although of fourth order in $V$, it is quadratic in both $D$ and $J$,
\begin{align}
	t_2 \sim \mathrm{e}^{\mathrm{i}( \varphi_{ij}-\varphi_{kj} )} \frac{(J D)^2}{B^3}.
	\label{eq:Haldane-type}
\end{align}
Importantly, two different bond directions and, hence, two different DMI angles, $\varphi_{ij}$ and $\varphi_{kj}$, are involved, rendering $t_2$ complex. Hence, $t_2$ resembles the complex Haldane-type second-nearest neighbor hopping that breaks TRS and opens topologically nontrivial band gaps \cite{Haldane1988}. Topological chiral edge states are expected, in agreement with Ref.~\onlinecite{mook2021}.
The predicted scaling of the magnon band gap with $D^{2}$ and $B^{-3}$ in Eq.~\eqref{eq:Haldane-type} agrees with the results of nonlinear spin-wave theory in Figs.~\ref{fig:FM}(a,b).


\section{Numerical simulations: DMRG and time evolution}

We complement our study of the (anti-)ferromagnetic honeycomb model with numerical methods such as Density Matrix Renormalization Group (DMRG) \cite{white_density_1992,mcculloch_infinite_2008,phien_infinite_2012} and real-time evolution.
The quantum many-body wave function is encoded as a matrix product state (MPS) with finite (auxiliary) bond dimension, and DMRG optimizes the entries of the MPS with respect to the energy.
In doing so, we obtain an MPS representation of the ground state.
Initially developed for one-dimensional systems, MPS and infinite DMRG (iDMRG) have been proven to be fairly unbiased and well controlled even for two-dimensional systems.
The extension to two dimensions is achieved by wrapping the lattice on a cylinder and winding the one-dimensional MPS structure along the cylinder.
Translational symmetry enables to treat infinitely long cylinders while the circumference remains finite, resulting in lines of accessible momenta in reciprocal space, see inset of Fig.~2(a) in the main text. 

In having obtained the ground state in MPS form,
we can now proceed to compute dynamical properties. 
We consider 
    $\mathcal S(\vec{k},\omega) = \sum_{\gamma \in \{x,y,z\}} \mathcal S^{\gamma\gamma} (\vec{k},\omega)$,
where   
$\mathcal S^{\gamma\gamma} (\vec{k},\omega)$
is the spatio-temporal Fourier transform of the dynamical correlations 
\begin{equation}
    \mathcal S^{\gamma\gamma} (\vec{k},\omega) = N \int \mathrm{d}t ~ \mathrm{e}^{\mathrm{i} \omega t} \sum_{a,b} \mathrm{e}^{\mathrm{i} \vec{k} \cdot (\vec{r}_b - \vec{r}_a)} ~ C^{\gamma\gamma}_{ab}(t) ~.
\end{equation}
Here, $\gamma \in {x,y,z}$ is the spin component,
$\vec{r}_a$ and $\vec{r}_b$ are the spatial positions of the spins,
and $N$ is a normalization factor.
$C^{\gamma \gamma}_{ab} (t)$ denotes the dynamical spin-spin correlation,
\begin{equation}
    C_{ab}^{\gamma \gamma} (t) = \langle \psi_0 | S_a^\gamma U(t) S_b^\gamma | \psi_0 \rangle~,
    \label{eq:SM_dyn_corr}
\end{equation}
which provides the protocol for the numerical time-evolution: 
(1) the ground state MPS, $|\psi_0\rangle$, is extended to a sufficiently long cylinder segment,
(2) a spin operator, $S_b^\gamma$, is applied to a site in the middle of the segment,
(3) the state is time-evolved by applying the time-evolution operator, $U(t) = \mathrm{e}^{\mathrm{i}H t}$,
(4) the second spin operator, $S_a^\gamma$, is applied,
and (5) the overlap with the ground state is computed. 
For computing the time evolution, we discretize $U(t) = \left[U(dt)\right]^N$ and represent $U(dt)$ as a matrix product operator~\cite{zaletel_timeevolving_2015}.
Index $a$ iterates over the two distinct sites within the unit cell in the middle of the MPS segment,
while $b$ iterates over all sites of the cylinder.

The initial excitation at $r_0$ will spread out with a characteristic \emph{light-cone} with the front moving as $r = r_0 + v_f t$, where $v_f$ is a velocity related to the maximal group velocity of the magnons. Hence, the length of the cylinder is chosen such, that the light cone does not reach the boundaries up to the final time in order to minimize boundary effects.
The entanglement within the light cone increases, requiring a larger $\chi_\text{bond}$ to encode the state up to a given accuracy.
We cap the bond dimension, $\chi_\text{bond} \le 256$, in order to keep the computational time within reasonable bounds.  
The time series is then extended by a linear prediction \cite{yule_vii_1927,makhoul_linear_1975,barthel_spectral_2009} and multiplied with a Gaussian, $g(t) = \exp(-t^2/(2\sigma^2))$, to suppress ringing.
As a consequence, we obtain a Gaussian line broadening in $\mathcal S(\bm k, \omega)$.
In the following we will summarize the numerical findings and parameters used for each model separately.

\subsection{Honeycomb XXZ antiferromagnet with DMI}
\begin{figure}
    \centering
    \includegraphics[width=\linewidth]{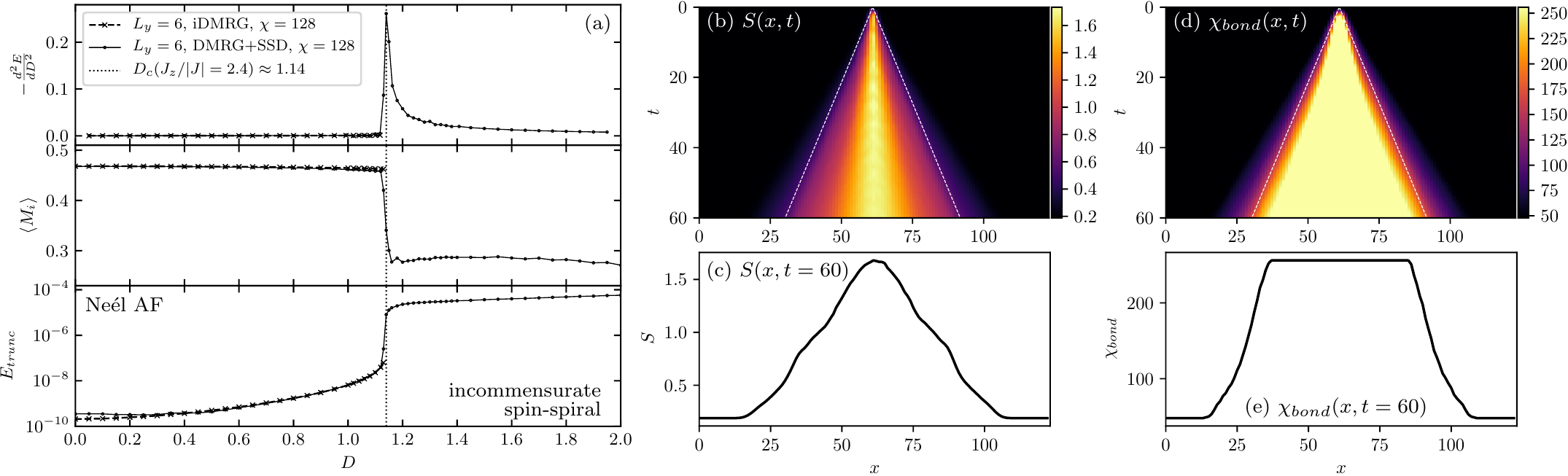}
    \caption{
        (a) Second derivative of the ground state energy, $d^2 E/dD^2$,
        local magnetic moments, $\langle M_i \rangle$,
        and the estimated error in the energy due to truncation, $E_\text{trunc}$,
        for the antiferromagnetic spin-$1/2$ honeycomb model with XXZ anisotropy of $J_z/|J|=2.4$ and varying DMI, $D$.
        For $D/|J| < D_c/|J| \approx 1.14$, the model is in a N\'eel state, while for $D > D_c$ the ground state is a spin-spiral with incommensurate wave vector, $q=q(D/|J|)$, depending on $D$.
        Finite DMRG on a cylinder with $L=(60,6)$ sites and \emph{sine-square deformation} (DMRG+SSD) is used to detect the spin-spiral phase, while infinite DMRG (iDMRG) only converges in the N\'eel phase.
        For $D/|J|<1.1$ a modest bond dimension of $\chi_\text{bond}=128$ is sufficient to reach a good accuracy of the MPS.
        (b) Entanglement entropy, $S(x,t)$, and (d) bond-dimension, $\chi_\text{bond}(x,t)$, during the time evolution of the excited MPS. 
        Both feature a typical \emph{light cone} structure originating from the excitation in the middle of the MPS segment. The dashed white line corresponds to the maximal group velocity extracted from the dispersion relation of the single magnon excitations.
        $\chi_\text{bond}$ is allowed to grow up to $\chi_\text{max}=256$ during the time evolution.
        (d,e) $t=60$ snapshots of $S(x,t)$ and $\chi_\text{bond}(x,t)$, respectively. 
     }
    \label{fig:SM_MPS_AF}
\end{figure}

The antiferromagnetic model [Eq.~(1) in the main text] exhibits a N\'eel ground state for small $D<D_c(J_z/J)$. While for sufficiently large $D$ the N\'eel state gives way to a spin-spiral with incommensurate wave vector.
Using a finite cylinder with a length of $L_{\parallel} = 60$ sites and a \emph{sine-square deformation} of the spin-exchange terms~\cite{gendiar_spherical_2009}, we find a phase transition at $D_c/J \approx 1.14$ for $J_z/J=2.4$ that appears to be continuous, see~Fig.~\ref{fig:SM_MPS_AF}(a).
iDMRG, on the other hand, enforces an ordering pattern that is commensurate with the iDMRG unit cell and is, thus, not well suited to capture the spin-spiral phase.
That is why we only include iDMRG data for $0 \le D < D_c$ in Fig.~\ref{fig:SM_MPS_AF}(a).

The relatively large XXZ anisotropy results in a large excitation gap
and reduces quantum fluctuations.
This is evident from the small reduction of the average local magnetic moments compared to the saturated value.
As a consequence, a modest bond dimension of $\chi_\text{bond} = 128$ is sufficient to encode the ground state within the N\'eel phase with high accuracy on cylinders with a circumference of $L_\text{circ}=6$ sites.

Here, we are only concerned with the magnon excitation in the N\'eel phase.
We compute $C_{ab}^{\gamma \gamma} (t)$ [see Eq.~\eqref{eq:SM_dyn_corr}]
using the MPS obtained from iDMRG. 
After extending it to $6 \times 123$ sites, 
the time-evolution is performed with time steps, $dt=0.01$,
while $C_{ab}^{\gamma \gamma} (t)$ is measured every $\delta t=0.2$ up to a total time of $t_\text{max} = 60$.
The bond dimension is capped at $\chi_\text{max} = 256$ during the time evolution.
Figures~\ref{fig:SM_MPS_AF}(b-e) show the evolution of the entanglement entropy, $S(x,t)$, for a bipartition at $x$, and the bond dimension $\chi(x,t)$.
After performing the Fourier transform in space, the time series $C^{\gamma \gamma} (\bm k, t)$ is extended to $\tilde t_\text{max} = 600$ using linear prediction and multiplied by a Gaussian with $\sigma_t = 25.12$.
Lastly, the temporal Fourier transform is applied. 
As a consequence of the convolution with the Gaussian,
the final $\mathcal S(\bm k, \omega)$ has a broadening of the magnon bands of $\sigma_\omega=0.040$ ($\sigma_\text{FWHM} = 0.094$).

\subsection{Honeycomb ferromagnet with DMI and external magnetic field}
\label{sec:technicalitiesFM}
Following the same procedure as for the antiferromagnetic model,
we find a phase transition at $B_{c1}/|J|\approx 0.06$ and $B_{c2}/|J|\approx 0.08$ for $D/|J| = 0.6$. 
For $B > B_{c2}$ the ground state is the fully polarized state.
All spins are fully aligned with the magnetic field, resulting in a product state that can be represented by an MPS with $\chi=1$.
The ground state has been proposed to be a spin-spiral for small $B$, and a skyrmion lattice at intermediate fields~\cite{muehlbauer_skyrmion_2009,han_skyrmion_2010} with the critical fields being $B_{c1} = 0.2\frac{D^2}{J}$ and $B_{c2} = 0.8\frac{D^2}{J}$, respectively~\cite{han_skyrmion_2010}.
While the sequence of phase transitions seems to agree with our numerical results,
we do not expect a quantitative agreement in the critical fields,
due to the restricting geometry used in our iDMRG simulation.

Here, we focus only on the magnon excitations in the ferromagetic phase at sufficiently large field.
Since the initial state is a product state, the computational cost of the time evolutions is reduced and we obtain $C_{ab}^{\gamma \gamma} (t)$ up to a time of $t_\text{max}=120$ on cylinders with $6 \times 129$ sites. 
The time-evolution is performed using time steps, $dt=0.01$,
while $C_{ab}^{\gamma \gamma} (t)$ is measured every $\delta t=0.2$.
The bond dimension is capped at $\chi_\text{max} = 256$ during the time evolution.
After performing the Fourier transform in space, the time series $C^{\gamma \gamma} (\bm k, t)$ is extended to $\tilde t_\text{max} = 1200$ using linear prediction and multiplied by a Gaussian with $\sigma_t = 58.60$.
This results in a broadening of $\sigma_\omega=0.017$ ($\sigma_\text{FWHM} = 0.040$) in $\mathcal S(\bm k, \omega)$.

\begin{widetext}

\section{Tetragonal model and general symmetry framework}

\begin{figure}
    \centering
    \includegraphics[width=0.5\linewidth,clip,trim={4.5 7.2cm 4.5 12.5}]{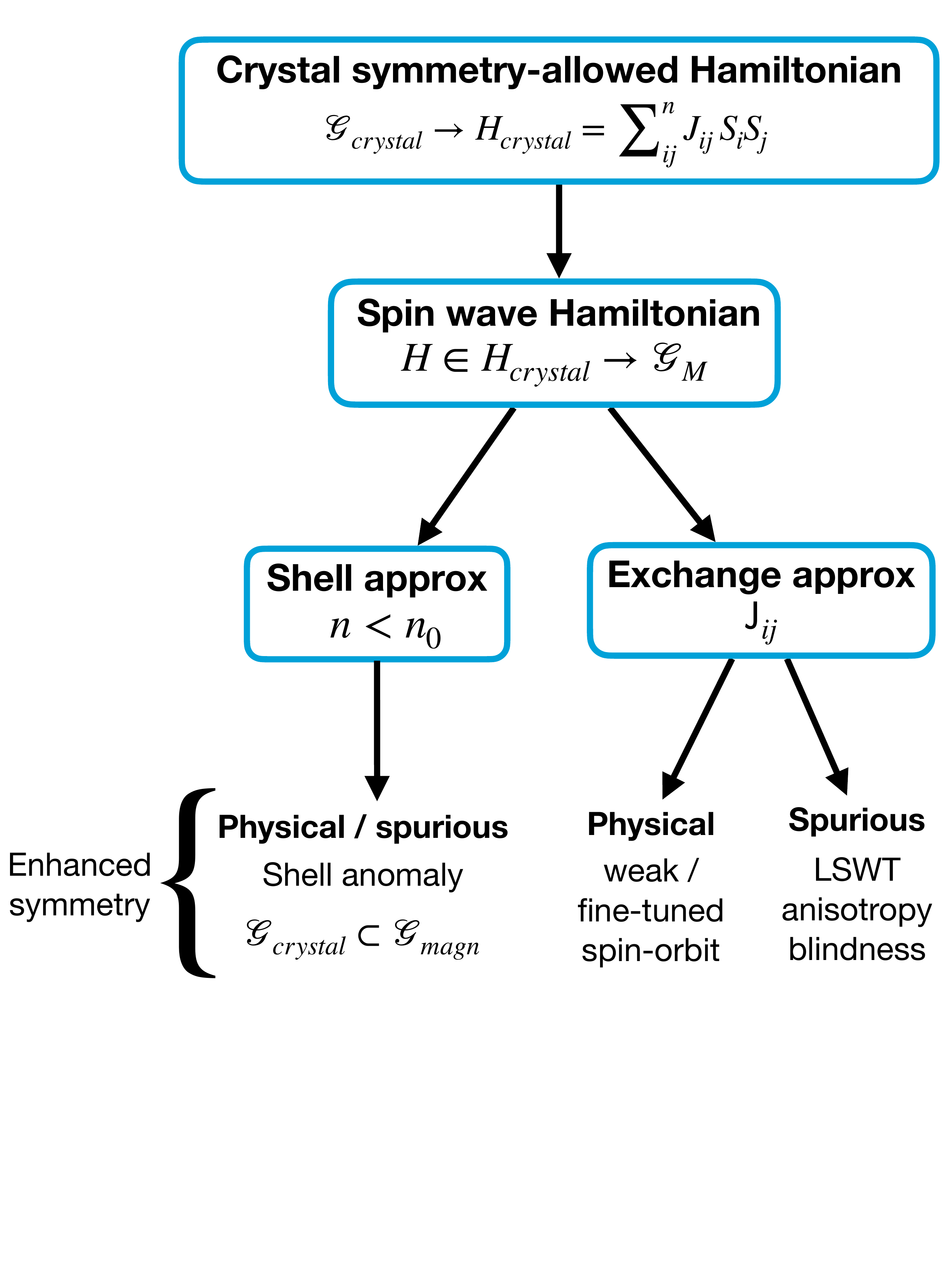}
    \caption{Scheme showing possible symmetry enhancing mechanisms in spin wave theory as explained in the text. The spin wave Hamiltonian $H$ must be an instance of the symmetry-allowed spin Hamiltonian of the crystal $H_{\text crystal}$. Approximations on interaction shell range $n<n_0$ and/or anisotropic exchange terms $\mathsf{J}_{ij}$ can lead to both spurious and physical enhanced symmetry in $\mathcal{G}_M$.}
    \label{fig:Enhanced_scheme}
\end{figure}

The spurious symmetry mechanisms given in the paper can be framed in a more general symmetry context as shown in Fig.~\ref{fig:Enhanced_scheme}.  
In general,  considering a crystal structure formed by both magnetic and non-magnetic ions,  the effective exchange Hamiltonian $H$ defined on the magnetic-only lattice has nevertheless embedded the symmetry of the whole structure.  This fact is due microscopically to the presence of spin-orbit coupling and of nonequivalent exchange paths mediated by non-magnetic ions. The first enforces the crystal symmetries by projecting anisotropic exchange couplings in the effective Hamiltonian, while the latter 
differentiates exchange couplings in the same shell of interaction. 

In such general case,  the effective exchange Hamiltonian group (paramagnetic group) is $\mathcal{G}_H= \mathcal{G}_{\text{crystal}} \oplus  \mathcal{T}\mathcal{G}_{\text{crystal}} $ where $\mathcal{G}_{\text{crystal}} $ is the space group of the whole crystal.  Note that time-reversal $ \mathcal{T}$ is preserved in the paramagnetic phase in absence of an external field.  Once the magnetic order is taken into consideration, pure $ \mathcal{T}$ is broken and in general, we reduce to a subgroup $\mathcal{G}_M \subset \mathcal{G}_H$ which is a magnetic space group.
Whatever additional symmetry in $\mathcal{G}_H$, or $\mathcal{G}_M$, with respect to this case is regarded as an enhanced symmetry.  Such enhanced symmetry can stem from a legitimate physical simplification or be an artifact derived by an incautious approximation. Here we treat them on the same basis, and only later we focus on the two spurious mechanisms, incautious choice of interaction shells and LSWT approximation.

Let us consider a magnetic lattice embedded in a crystal where spin-orbit coupling is present. We can presume symmetry is locked to transform the same in spin and real space and the symmetries of the whole structure are the relevant ones.  We can then derive the symmetry-allowed exchange Hamiltonian $H_{\text{crystal}}$ which is the most generic exchange Hamiltonian abiding by such lattice symmetries.
We limit ourselves to a bilinear interaction between moments:
\begin{align}
H_{\text{crystal}} = \sum_{i,j}^{n \rightarrow \infty} J_{ij}^{\,\mu\nu} S^\mu_i  S^\nu_j \,,
\label{Eq:ExchangeHam}
\end{align}
where all the interaction shells $n$ are considered.
In a totally generic exchange Hamiltonian,  the exchange matrix for a specific bond $ij$ may have nine allowed independent couplings:
\begin{align}
\mathsf{J}_{ij} =
\begin{pmatrix} 
\hphantom{-}J^{xx}     & \hphantom{-}J^{xy}       &  \hphantom{-}J^{xz}     \\
\hphantom{-}J^{yx}     & \hphantom{-}J^{yy}       &  \hphantom{-}J^{yz}     \\
\hphantom{-}J^{zx}     & \hphantom{-}J^{zy}        & \hphantom{-}J^{zz} 
\end{pmatrix} 
\end{align}
and each bond $ij$ is totally independent of another.
However the symmetry of the crystal $g \in \mathcal{G}_{\text{crystal}}$ will enforce conditions $J_{i'j'} = [g_{\mu\nu} \,J_{ij}^{\,\mu\nu}]$, reducing the number of independent couplings and relating different bonds $\langle ij \rangle \, \xrightarrow{g} \, \langle i'j' \rangle$, producing the symmetry-allowed exchange Hamiltonian $H_{\text{crystal}}$ for $\mathcal{G}_{\text{crystal}}$ at the top of Fig. \ref{fig:Enhanced_scheme} (as example see the tetragonal model \ref{subsubsec:TetragonalHam}).

The effective spin wave Hamiltonian $H$ will be a simplified instance of this generic symmetry-allowed Hamiltonian $H_{\text{crystal}}$ depending on the microscopic detail of the system.
For example, such effective Hamiltonian will have generally only a few interaction shells $n<n_0$ over which the microscopic exchange paths,  and therefore the couplings,  are significant.
We can call the enhanced symmetry produced in such a way {\it physical shell anomaly}.
Also, the degree of spin-orbit may lead to fewer anisotropic exchange terms in $J^{\,\mu\nu}$ for a given shell with respect to the symmetry-allowed ones. Such physical simplifications can lead to enhanced symmetry and even possibly unlocking spin and real space transformations. 
This decoupling is formally described by a spin-space group \cite{BrinkmanElliott1966} which tends to lead to magnon spectra with more degeneracies than the original magnetic space group \cite{corticelli2022}.
An element of a spin-space group can be written as $\sselement{B}{R}{\bs{t}}$ where $B$ acts on the spin and $\{R|\bs{t}\}$ on the real space independently \cite{Litvin1974}. The operator $B$ is a spin rotation, while $\{R|\bs{t}\}$ is a real space rotation $R$ followed by a translation $\bs{t}$.
Following this notation, an element of the magnetic space group would be of the form $\sselement{R}{R}{\bs{t}}$.
Spin-space groups can be seen as supergroups of magnetic space groups and  they are the most general symmetry framework to study symmetry in the magnetic materials and therefore will be used in this treatment.
There are many physically well-motivated cases where the exchange couplings have spin-space symmetry---for example in Heisenberg, $XY$, Kitaev or Dzyaloshinskii–Moriya models---generally when spin-orbit is weak or fine-tuned and there are only a few relevant exchange couplings. 

On top of the physical microscopic simplification, we usually approximate the effective exchange Hamiltonian to compute the spin wave spectrum. Many times indeed, we are interested in a qualitative agreement, more than a quantitative one.   We could,  for example,  cut further the interaction shells for simplicity (potentially producing a {\it spurious shell anomaly}).  Again,  we could use a linear spin wave approximation, which is effectively an approximation on the exchange terms (or {\it anisotropy blindness}).  We could incautiously think that we will obtain only a quantitative change and indeed this is often the case. Nonetheless, with such approximations, it is also possible to introduce new enhanced symmetry which changes qualitative the physics of the system like the spectrum degeneracy and its related magnon band topology.

A systematic check for extra symmetries, both physical and spurious,  is possible in the following way.
For an extra symmetry $\sselement{B}{R}{\bs{t}}$ to be present, we have conditions that must simultaneously met for both $\{R|\bs{t}\}$ and $B$. The real space transformation $\{R|\bs{t}\}$ must respect the magnetic ions lattice symmetries on top of which $H$ is defined.
On the other hand, the spin rotation $B$ must respect the relations between interacting bonds (exchange matrices) of the magnetic Hamiltonian.
Once the magnetic order is considered, the spin rotation $B$ must also respect it.

The easier thing to check is if the exchange relations allow free spin rotations $\sselement{B}{E}{\bs{0}}$ in the effective Hamiltonian ($E$ is the identity transformation). 
Afterwards, we can look for elements with a real space non-trivial transformation.
As we said the condition on $\{R|\bs{t}\}$ is met by all the symmetries of the underlying magnetic ions lattice, which we call $ \mathcal{G}_{\text{magn}}$. The magnetic ions are only a part of the whole crystal, therefore it is always true that $ \mathcal{G}_{\text{crystal}} \subseteq \mathcal{G}_{\text{magn}}$.
The interesting $\{R|\bs{t}\}$ to check are the ones "latent" for the magnetic structure, so present in $\mathcal{G}_{\text{magn}}$ but not in $ \mathcal{G}_{\text{crystal}}$. The more latent symmetries there are,  the easier is to have enhanced symmetries from shell anomaly.
Once a proper $\{R|\bs{t}\}$ is selected, it must be paired with a $B$ which respects exchange symmetry relations. If this pairing is possible we have an enhanced symmetry in $\mathcal{G}_H$, and in case $B \neq R$, it is a spin-space symmetry.  In case a magnetic order is present,  we should check that all such new elements respect it,  enhancing therefore $\mathcal{G}_M$.

{\it LSWT symmetry enhancement.}  
From a symmetry group perspective the blindness to longitudinal-transverse components (anisotropy blindness) of the linear spin wave theory imposes a further constraint on the exchange for each $\langle ij \rangle$:
\begin{align}
\mathsf{J}_{ij} =
\begin{pmatrix} 
\hphantom{-}J^{xx}     & \hphantom{-}J^{xy}       &  \hphantom{-}0     \\
\hphantom{-}J^{yx}     & \hphantom{-}J^{yy}       &  \hphantom{-}0     \\
\hphantom{-}0     & \hphantom{-}0       & \hphantom{-}J^{zz} 
\end{pmatrix} 
\label{eq:LSWTcondition}
\end{align}
which corresponds to an extra $C_2$ rotation of the spins around their local quantization axis $z$. 
Indeed the terms $J^{\alpha z}$ and $J^{z \alpha}$ with $\alpha = x,y$ do not enter into the quadratic expansion.
The LSWT spectrum will therefore abide by this extra constraint and possibly show extra symmetries in its relevant group  $\mathcal{G}_{\rm LSWT} \supseteq \mathcal{G}_M$.  More specifically, the LSWT group $\mathcal{G}_{\rm LSWT}$ can be one of  $\{\mathcal{G}_M, \mathcal{G}^*_M, \mathcal{G}_E, \mathcal{G}^*_E \}$ as shown in Fig.~$1$ in the main text.

For a {\it collinear} system, $\mathcal{G}_{\rm LSWT} = \{\mathcal{G}^*_M, \mathcal{G}^*_E \}$ since it includes always a spurious global spin rotation $\sselement{2_{001}}{E}{0}$ around the collinear axis $[001]$, which nevertheless has usually a trivial effect on the spectrum. However, in model B in the main text, such spurious global spin rotation switch off the DMI term, allowing free rotation U(1) of the spins around the Ising axis and enhancing the group to $\mathcal{G}_M^*$.
More subtle in general, are the enhanced symmetries $\mathcal{G}_{\rm LSWT} = \mathcal{G}^*_E$ coming from an interplay with shell anomaly, where the latent symmetries of the magnetic ions lattice $\mathcal{G}_{\text{magn}}$ become relevant.
Here indeed, the LSWT exchange Hamiltonian has an important extra constraint Eq.~\eqref{eq:LSWTcondition}, allowing for easier matching of $B$ and $\{R|\bs{t}\}$ and therefore additional symmetries with possible spurious degeneracies.  We will see an example of this mechanism in tetragonal model C in \ref{subsubsec:lswenhancementTetragonal}.

\subsection{Tetragonal Lattice model}
\label{sec:tet-model}

\subsubsection{Lattice detail}

In this section we spell out details of the model C in the main text. This is based on ferromagnetic tetragonal crystal structure with space group P4 (\#75) and magnetic moments aligned in the $z$ direction $[001]$.
The primitive vectors spanning a tetragonal lattice are:
\begin{align}
    \bs{a}_1 = (a,0,0),~~ \bs{a}_2 = (0,a,0), ~~\bs{a}_3 = (0,0,c)
\end{align}
The basis where the spins sit is (Wyckoff postion $2c$):
\begin{align}
\bs{d}_1 = (0,1/2,z),~~ \bs{d}_2 = (1/2,0,z)
\end{align}
There are 4 (axial) point group symmetries forming the space group:
\begin{align}
\mathcal{G}_{\text crystal} =  &\element{E}{{\bf 0}}, ~ \element{2_{001}}{{\bf 0}}, ~
\{4^+_{001}|\bs{0}\}, ~ \{4^-_{001}|\bs{0}\}, \\
&\element{E}{{1,0,0}}, \element{E}{{0,1,0}}, \element{E}{{0,0,1}}
\end{align}
where the second line are the primitive lattice translations.
Also we define for later use in the exchange coupling matrix the additional neighboring lattice points:
\begin{align}
\bs{d}_{1x} = \bs{d}_1 +  \bs{a}_1 ,~~~ \bs{d}_{1y} = \bs{d}_1 +  \bs{a}_2, ~~~\bs{d}_{1z} = \bs{d}_1 +  \bs{a}_3  \nonumber \\
\bs{d}_{2x} = \bs{d}_2 +  \bs{a}_1 ,~~~ \bs{d}_{2y} = \bs{d}_2 +  \bs{a}_2, ~~~\bs{d}_{2z} = \bs{d}_2 +  \bs{a}_3
\end{align}

\begin{figure}
    \centering
    \includegraphics[width=0.26\linewidth,clip,trim={4.5 1.2cm 4.5 12.5}]{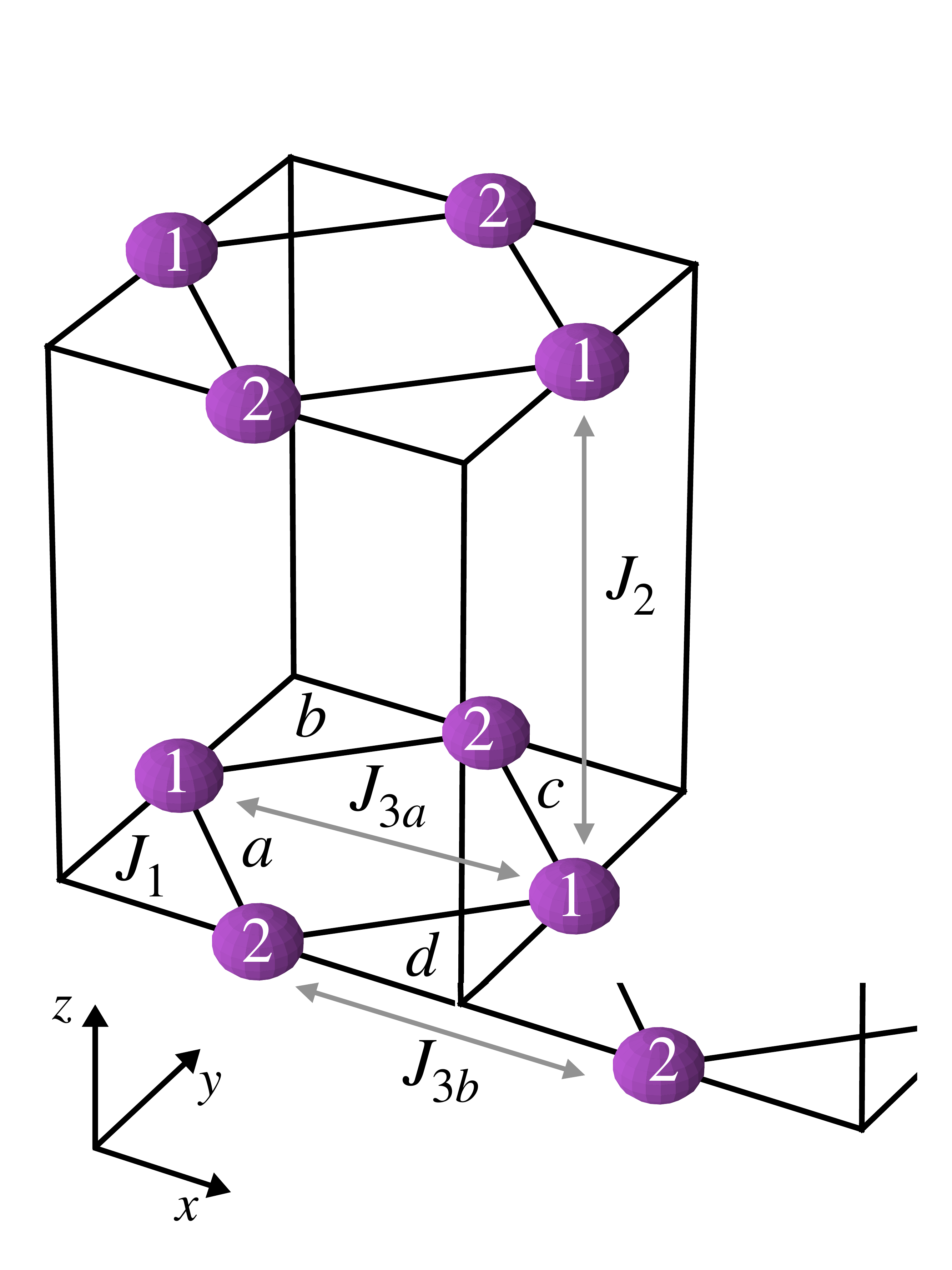}
    \caption{The figure shows the tetragonal lattice of space group P4 (\#75) (model C in the main text) with the relevant bonds up to third shell.}
    \label{fig:Tetragonal_model_supplementary}
\end{figure}

\subsubsection{Symmetry-allowed exchange Hamiltonian}
\label{subsubsec:TetragonalHam}

Here we list all the possible couplings which are allowed by the symmetry of $\mathcal{G}_{\text crystal}$ up to the third shell.
The minimal model to obtain a 3D connected lattice has two shells: the $J_1$ and $J_2$ bonds. 
The third shell has two independent kinds of bond $J_{3a}$ and $J_{3b}$.
Considering all the three shells together we have a total of 10 bonds, divided into 4 independent equivalence sets. A set of equivalent bonds contains bonds that can be transformed into each other by symmetry operations of $\mathcal{G}_{\text crystal}$.
The bonds are, referencing the tetragonal lattice in Fig.~\ref{fig:Tetragonal_model_supplementary}:
\begin{itemize}
    \item Set $J_1$ : ${(1,2)_a,(1,2)_b,(1,2)_c,(1,2)_d}$
	\item Set $J_2$ : ${(1,1)_z,(2,2)_z}$
    \item Set $J_{3a}$ : ${(1,1)_x,(2,2)_y}$
    \item Set $J_{3b}$ : ${(1,1)_y,(2,2)_x}$
\end{itemize}

Their symmetry-allowed exchanges read:
\begin{align}
\hspace{0px}
\mathsf{J}_{(1, 2)_a} = & 
\begin{pmatrix} 
\hphantom{-}J_1^{xx}     & \hphantom{-}J_1^{xy}       &  \hphantom{-}J_1^{xz}     \\
\hphantom{-}J_1^{yx}     & \hphantom{-}J_1^{yy}       &  \hphantom{-}J_1^{yz}     \\
\hphantom{-}J_1^{zx}     & \hphantom{-}J_1^{zy}        & \hphantom{-}J_1^{zz} 
\end{pmatrix} 
\hspace{40px}&
\mathsf{J}_{(1, 2)_b}  = &
\begin{pmatrix} 
\hphantom{-}J_1^{yy}     & -J_1^{xy}       &  \hphantom{-}J_1^{zy}     \\
-J_1^{yx}                              & \hphantom{-}J_1^{xx}       &  -J_1^{zx}     \\
\hphantom{-}J_1^{yz}     & -J_1^{xz}        & \hphantom{-}J_1^{zz} 
\end{pmatrix}
\label{Eq:Jcoupling12a}
\\[6px]
\mathsf{J}_{(1, 2)_c}  = &
\begin{pmatrix} 
\hphantom{-}J_1^{xx}     & \hphantom{-}J_1^{xy}       &  -J_1^{xz}     \\
\hphantom{-}J_1^{yx}     & \hphantom{-}J_1^{yy}       &  -J_1^{yz}     \\
-J_1^{zx}     & -J_1^{zy}        & \hphantom{-}J_1^{zz} 
\end{pmatrix}
&
\mathsf{J}_{(1, 2)_d}  = &
\begin{pmatrix} 
\hphantom{-}J_1^{yy}     & -J_1^{xy}       &  -J_1^{zy}     \\
-J_1^{yx}                              & \hphantom{-}J_1^{xx}       &  \hphantom{-}J_1^{zx}     \\
-J_1^{yz}     & \hphantom{-}J_1^{xz}        & \hphantom{-}J_1^{zz} 
\end{pmatrix}
  \\[6px]
\mathsf{J}_{(1, 1)_z} = & 
\begin{pmatrix} 
\hphantom{-}J_2^{xx}     & \hphantom{-}J_2^{xy}       &  \hphantom{-}0     \\
\hphantom{-}J_2^{yx}     & \hphantom{-}J_2^{yy}       &  \hphantom{-}0    \\
\hphantom{-}0     &\hphantom{-}0        & \hphantom{-}J_2^{zz} 
\end{pmatrix} 
&
\mathsf{J}_{(2, 2)_z} = & 
\begin{pmatrix} 
\hphantom{-}J_2^{yy}     & -J_2^{yx}       &  \hphantom{-}0     \\
-J_2^{xy}     & \hphantom{-}J_2^{xx}       &  \hphantom{-}0    \\
\hphantom{-}0     &\hphantom{-}0        & \hphantom{-}J_2^{zz} 
\end{pmatrix} 
 \\[6px]
\mathsf{J}_{(1, 1)_x} = & 
\begin{pmatrix} 
\hphantom{-}J_{3a}^{xx}     & \hphantom{-}J_{3a}^{xy}       &  \hphantom{-}J_{3a}^{xz}     \\
\hphantom{-}J_{3a}^{xy}     & \hphantom{-}J_{3a}^{yy}       &  \hphantom{-}J_{3a}^{yz}     \\
-J_{3a}^{xz}     & -J_{3a}^{yz}        & \hphantom{-}J_{3a}^{zz} 
\end{pmatrix} 
\hspace{40px}&
\mathsf{J}_{(2, 2)_y}  = &
\begin{pmatrix} 
\hphantom{-}J_{3a}^{yy}     & -J_{3a}^{xy}       &  -J_{3a}^{yz}     \\
-J_{3a}^{xy}     & \hphantom{-}J_{3a}^{xx}       &  \hphantom{-}J_{3a}^{xz}     \\
\hphantom{-}J_{3a}^{yz}     & -J_{3a}^{xz}        & \hphantom{-}J_{3a}^{zz} 
\end{pmatrix}
 \\[6px]
\mathsf{J}_{(1, 1)_y} = & 
\begin{pmatrix} 
\hphantom{-}J_{3b}^{xx}     & \hphantom{-}J_{3b}^{xy}       &  \hphantom{-}J_{3b}^{xz}     \\
\hphantom{-}J_{3b}^{xy}     & \hphantom{-}J_{3b}^{yy}       &  \hphantom{-}J_{3b}^{yz}     \\
-J_{3b}^{xz}     & -J_{3b}^{yz}        & \hphantom{-}J_{3b}^{zz} 
\end{pmatrix} 
\hspace{40px}&
\mathsf{J}_{(2, 2)_x}  = &
\begin{pmatrix} 
\hphantom{-}J_{3b}^{yy}     & -J_{3b}^{xy}       &  \hphantom{-}J_{3b}^{yz}     \\
-J_{3b}^{xy}     & \hphantom{-}J_{3b}^{xx}       &  -J_{3b}^{xz}     \\
-J_{3b}^{yz}     & \hphantom{-}J_{3b}^{xz}        & \hphantom{-}J_{3b}^{zz} 
\end{pmatrix}
\end{align}
Therefore for $J_1+J_2$ ($J_1 +J_2 +J_3$) we have 14 (26) possible couplings which reduce to 10 (18) if we apply the LSWT constraints.
We list here also the specific parameters used in the main text for the $J_1 +J_2 +J_3$ model (chosen to be maximally anisotropic):
\begin{align}
\hspace{0px}
\mathsf{J}_{(1, 2)_a} = & 
\begin{pmatrix} 
-0.445                & \hphantom{-}0.174     &  \hphantom{-}0.260     \\
-0.475                & -0.221                &  \hphantom{-}0.322     \\
\hphantom{-}0.028     & \hphantom{-}0.049     & \hphantom{-}0.028 
\end{pmatrix} 
\hspace{40px}&
\mathsf{J}_{(1, 1)_z} = & 
\begin{pmatrix} 
\hphantom{-}0.358   & -0.182         &  \hphantom{-}0     \\
-0.083              & -0.396         &  \hphantom{-}0    \\
\hphantom{-}0       &\hphantom{-}0   & 0.469
\end{pmatrix} 
\\[6px]
\mathsf{J}_{(1, 1)_x} = & 
\begin{pmatrix} 
-0.093    & \hphantom{-}0.084     &  \hphantom{-}0.0009    \\
0.084    & -0.0008               &  -0.097                \\
-0.0009   & \hphantom{-}0.097     & -0.035 
\end{pmatrix}
&
\mathsf{J}_{(1, 1)_y} = & 
\begin{pmatrix} 
\hphantom{-}0.083    & \hphantom{-}0.070   &  \hphantom{-}0.085    \\
0.070               & -0.018              &  0.0003               \\
-0.085               & -0.0003             & \hphantom{-}0.045 
\end{pmatrix} 
\end{align}

\subsubsection{Linear Spin Wave Hamiltonian}

We consider the presence of a magnetic field with a magnitude $\mathcal{B}$\footnote{We chose $\mathcal{B}$ instead of $B$ to denote the magnetic field to avoid confusion with the spin rotation $B$.} greater than some threshold in such a way that the state is fully polarized in the field direction. In this situation we can expand the moments in small fluctuations about this collinear state in Holstein-Primakoff bosons. 
Here we consider a field direction along $[001]$ so the local frame is the same as the laboratory frame and no local rotation is needed. The Hamiltonian in the local frame reads:
\begin{align}
H =    \sum_{<ij>} \sum_{\alpha,\beta=x,y,z} \bs{S}_i^{\alpha}\, \mathsf{J}^{\alpha\beta}\, \bs{S}_j^{\beta} - \mathcal{B} \sum_i \bs{S}_i^z
\end{align}
The linear spin wave Hamiltonian is
\begin{align}
H_{LSWT} =\frac{S}{2} \sum_{\bs{k}}  \Upsilon_{\bs{k}}^{\dagger}\bs{M}(\bs{k}) \Upsilon_{\bs{k}}
\end{align}
where
\begin{align}
 \Upsilon_{\bs{k}} = \big(b_{1,\bs{k}},b_{2,\bs{k}},b_{1,-\bs{k}}^{\dagger},b_{2,-\bs{k}}^{\dagger}\big)^\text{T}
\end{align}
with $b_i$ and $b_i^{\dagger}$ bosonic operators living on the sublattice generated by the basis $\bs{d}_i$.
The $4\times4$ matrix $\bs{M}(\bs{k})$ takes then the form
\begin{align}
\bs{M}(\bs{k}) = 
\left(
\begin{matrix}
\mathsf{A}(\bs{k}) & \mathsf{B}(\bs{k}) \\
\mathsf{B}^{*}(-\bs{k}) & \mathsf{A}^*(-\bs{k})
\end{matrix}
\right)
\end{align}
where the $\mathsf{A}_{ab}(\boldsymbol{k})$ and $\mathsf{B}_{ab}(\boldsymbol{k})$ depend on the exchange couplings in the local frame as follows:
\begin{align}
\mathsf{A}_{ab}(\boldsymbol{k}) & =  \frac{1}{2}\left(\mathsf{J}_{ab}^{xx}(\boldsymbol{k}) + \mathsf{J}_{ab}^{yy}(\boldsymbol{k}) -\mathrm{i} \mathsf{J}_{ab}^{xy}(\boldsymbol{k}) + \mathrm{i}\mathsf{J}_{ab}^{yx}(\boldsymbol{k})   \right) - \delta_{ab} \sum_{c} (\mathsf{J}_{ac}^{zz}(\boldsymbol{0}) - \mathcal{B}/S)~, \label{eq:Ablock} \\
\mathsf{B}_{ab}(\boldsymbol{k}) & = \frac{1}{2}\left(\mathsf{J}_{ab}^{xx}(\boldsymbol{k}) - \mathsf{J}_{ab}^{yy}(\boldsymbol{k}) + \mathrm{i} \mathsf{J}_{ab}^{xy}(\boldsymbol{k}) + \mathrm{i}\mathsf{J}_{ab}^{yx}(\boldsymbol{k})   \right)~.
\label{eq:Bblock}
\end{align}
Explicitly the Fourier transformed exchange couplings for the tetragonal model are:
\begin{align}
\mathsf{J}_{11} = ~ & ~~~\mathsf{J}_{(1,1)_z} \Exp{\mathrm{i}\bs{k}\cdot(\bs{d}_1-\bs{d}_{1z})}  +  \mathsf{J}_{(1,1)_x} \Exp{\mathrm{i}\bs{k}\cdot(\bs{d}_1-\bs{d}_{1x})}  +  \mathsf{J}_{(1,1)_y} \Exp{\mathrm{i}\bs{k}\cdot(\bs{d}_1-\bs{d}_{1y})} \nonumber \\
&+ \mathsf{J}_{(1,1)_z}^T \Exp{\mathrm{i}\bs{k}\cdot(\bs{d}_{1z} -\bs{d}_{1})}  +  \mathsf{J}_{(1,1)_x}^T \Exp{\mathrm{i}\bs{k}\cdot(\bs{d}_{1x}-\bs{d}_{1})}  +  \mathsf{J}_{(1,1)_y}^T \Exp{\mathrm{i}\bs{k}\cdot(\bs{d}_{1y}-\bs{d}_{1})}~,
 \\
\mathsf{J}_{22} = ~ & ~~~\mathsf{J}_{(2,2)_z} \Exp{\mathrm{i}\bs{k}\cdot(\bs{d}_2-\bs{d}_{2z})}  +  \mathsf{J}_{(2,2)_x} \Exp{\mathrm{i}\bs{k}\cdot(\bs{d}_2-\bs{d}_{2x})}  +  \mathsf{J}_{(2,2)_y} \Exp{\mathrm{i}\bs{k}\cdot(\bs{d}_2-\bs{d}_{2y})}  \nonumber \\
&+ \mathsf{J}_{(2,2)_z}^T \Exp{\mathrm{i}\bs{k}\cdot(\bs{d}_{2z} -\bs{d}_{2})}  +  \mathsf{J}_{(2,2)_x}^T \Exp{\mathrm{i}\bs{k}\cdot(\bs{d}_{2x}-\bs{d}_{2})}  +  \mathsf{J}_{(2,2)_y}^T \Exp{\mathrm{i}\bs{k}\cdot(\bs{d}_{2y}-\bs{d}_{2})}~,
 \\
\mathsf{J}_{12} = ~ & ~~~\mathsf{J}_{(1,2)_a} \Exp{\mathrm{i}\bs{k}\cdot(\bs{d}_1-\bs{d}_{2})}  +  \mathsf{J}_{(1,2)_b} \Exp{\mathrm{i}\bs{k}\cdot(\bs{d}_1-\bs{d}_{2y})}  +  \mathsf{J}_{(1,2)_c} \Exp{\mathrm{i}\bs{k}\cdot(\bs{d}_{1x}-\bs{d}_{2y})} +  \mathsf{J}_{(1,2)_d} \Exp{\mathrm{i}\bs{k}\cdot(\bs{d}_{1x}-\bs{d}_{2})}~,
 \\
\mathsf{J}_{21} = ~ & ~~~\mathsf{J}_{(1,2)_a}^T \Exp{\mathrm{i}\bs{k}\cdot(\bs{d}_2-\bs{d}_{1})}  +  \mathsf{J}_{(1,2)_b}^T \Exp{\mathrm{i}\bs{k}\cdot(\bs{d}_{2y}-\bs{d}_{1})}  +  \mathsf{J}_{(1,2)_c}^T \Exp{\mathrm{i}\bs{k}\cdot(\bs{d}_{2y}-\bs{d}_{1x})} +  \mathsf{J}_{(1,2)_d}^T \Exp{\mathrm{i}\bs{k}\cdot(\bs{d}_{2}-\bs{d}_{1x})}~.
\end{align}

\subsubsection{LSWT symmetry enhancement}
\label{subsubsec:lswenhancementTetragonal}

Since here the model is more involved, it is useful to find a more systematic way to scan possible enhanced degeneracy. We exploit therefore the general symmetry framework based on spin-space groups discussed above.
First, since the magnetism here is collinear, there is an extra pure spin rotation symmetry $\sselement{2_{001}}{E}{0}$ in the LSWT, regardless of the couplings, due to the anisotropy blindness. Therefore the LSWT group $\mathcal{G}_{\rm LSWT}$ must be $\mathcal{G}^*_M$ or $\mathcal{G}^*_E$.
Second, we want to explore the non-trivial symmetry of the underlying magnetic ions lattice $\{R|\bs{t}\}$ which coupled with a spin rotation $B$ can produce an extra spin-space symmetry $\sselement{B}{R}{\bs{t}}$.

Here the magnetic lattice is hosted on the Wyckoff 2c of a crystal (including non-magnetic ions) with group \#75 P4 of elements:
\begin{align}
\mathcal{G}_M =  &\element{E}{{\bf 0}}, ~ \element{2_{001}}{{\bf 0}}, ~ \{4^+_{001}|{\bf 0}\}, ~ \{4^-_{001}|{\bf 0}\}, \\
&\element{E}{{1,0,0}}, \element{E}{{0,1,0}}, \element{E}{{0,0,1}}
\end{align}
where we consider $\mathcal{G}_M = \mathcal{G}_{\text{crystal}}$, so the magnetic order respect the crystal symmetry.
However, the magnetic lattice alone forms a simple tetragonal Bravais lattice, with a corresponding space group \#123 P4/mmm:
\begin{align}
\mathcal{G}_{\rm magn} =  &\element{E}{{\bf 0}},  \element{2_{001}}{{\bf 0}},  \{4^{\pm}_{001}|\bs{0}\}, \element{2_{010}}{{\bf 0}}, \element{2_{100}}{{\bf 0}}, \element{2_{110}}{{\bf 0}}, \element{2_{1-10}}{{\bf 0}}, \\ \nonumber
                                              &\element{-1}{{\bf 0}}, \element{m_{001}}{{\bf 0}}, \{-4^{\pm}_{001}|\bs{0}\}, \element{m_{010}}{{\bf 0}}, \element{m_{100}}{{\bf 0}},  \element{m_{110}}{{\bf 0}}, \element{m_{1-10}}{{\bf 0}}, \\ \nonumber
                                              & \element{E}{{1/2,1/2,0}}, \element{E}{{0,1,0}}, \element{E}{{0,0,1}}\\ \nonumber
\end{align}
We note here that $\mathcal{G}_{M} = \mathcal{G}_{\text{crystal}} \subseteq \mathcal{G}_{\text{magn}}$, indeed the first four elements and two of the primitive lattice belongs to $\mathcal{G}_{M}$. The interesting symmetries to check are therefore the remaining latent symmetries, of which it is enough to check the representative set
\begin{align}
\mathbf{L} = \element{-1}{{\bf 0}},  \element{m_{100}}{{\bf 0}}, \element{E}{{1/2,1/2,0}}
\end{align}
and all the internal combinations (7 possible cases). 
These tentative real space symmetries must be paired now with a spin rotation $B$ in order to make a possible spin-space element $\sselement{B}{R}{\bs{t}}$ that preserves the magnetic order and the exchange symmetry constraints in \ref{subsubsec:TetragonalHam}, considering also the anisotropy blindness of LSWT.
For each shell of interaction we can do this calculation and the resulting representative enhancing elements are shown in Tab.~\ref{tab:LSWgroup}.
We can see that for the $J_1 + J_2$ model we have $\mathcal{G}_{\rm LSWT} = \mathcal{G}^*_E$, so a combination of anisotropy blindness and shell anomaly. Instead for $J_1 + J_2 + J_3$ we have only anisotropy blindness $\mathcal{G}_{\rm LSWT} = \mathcal{G}^*_M$, but no consequence on the spectrum, since the triviality of the spin rotation.

\begin{table*}[!t]
\centering
\hbox to \linewidth{ \hss
\begin{tabular}{ |c|p{9.0cm}|   }
 \hline
 \multicolumn{1}{|c|}{Couplings}  & Representatives spurious enhanced symmetry in $\mathcal{G}_{\rm LSWT}$ \\ 
 \hline
 $J_1 $        &  $\sselement{2_{001}}{E}{0} + \sselement{E}{-1}{0} + \sselement{4_{001}^{+}}{m_{100}}{1/2,1/2,0}$      \\
 $J_2 $        &  $\sselement{2_{001}}{E}{0} + \sselement{E}{m_{100}}{0} + \sselement{4_{001}^{+}}{E}{1/2,1/2,0}$      \\
 $J_3 $        &  $\sselement{2_{001}}{E}{0} + \sselement{E}{-1}{0} + \sselement{E}{m_{100}}{0}$      \\
$J_1 + J_2 $  &  $\sselement{2_{001}}{E}{0} + \sselement{4_{001}^{+}}{m_{100}}{1/2,1/2,0}$      \\
 $J_1 + J_3 $        &  $\sselement{2_{001}}{E}{0} + \sselement{E}{-1}{0}$      \\
 $J_2 + J_3$        &  $\sselement{2_{001}}{E}{0} + \sselement{E}{m_{100}}{0}$      \\
 $J_1 + J_2 + J_3 $ &  $\sselement{2_{001}}{E}{0}$      \\
 \hline
\end{tabular}
\hss}
\caption{
Spin-space representative symmetries for each bond coupling, enhancing the original space group $\mathcal{G}_{M}$. 
}
\label{tab:LSWgroup}
\end{table*}

As example of the calculation let us take the extra spin glide symmetry $\sselement{4_{001}^{+}}{m_{100}}{1/2,1/2,0}$ responsible for the degeneracy in the $J_1 + J_2$ model.
We focus on the shell $J_1$ and we operate with the glide on two of the 4 bonds of the set for simplicity:
\begin{align}
&(1,2)_a  \xrightarrow{\{m_{100}|1/2,1/2,0\}} (2,1)_b ~,\\  \nonumber
&(1,2)_b  \xrightarrow{\{m_{100}|1/2,1/2,0\}} (2,1)_a ~.
\end{align}
So upon applying the glide these two bonds swap. Originally the bonds $(1,2)_a$ and $(1,2)_b$ were related in the model by two counter-rotating $C_4$:
\begin{align}
&(1,2)_a  \xrightarrow{\sselement{4_{001}^{+}}{4_{001}^{+}}{\bf 0}} (2,1)_b ~, \\  \nonumber
&(1,2)_b  \xrightarrow{\sselement{4_{001}^{-}}{4_{001}^{-}}{\bf 0}} (2,1)_a ~,
\end{align}
and this is responsible for the exchange constraints (Eq. \ref{Eq:Jcoupling12a} etc.). 
To preserve this exchange matrix relation we may choose as the spin rotation $B$ a $C_4$ rotation which in combination with the glide gives $\sselement{4_{001}^{+}}{m_{100}}{1/2,1/2,0}$, but we immediately see that only the first line is satisfied while the second one would require $\sselement{4_{001}^{-}}{m_{100}}{1/2,1/2,0}$, a rotation in the opposite direction. These conditions cannot be satisfied simultaneously and the extra symmetry is not present.
However if we now allow for the presence of the $\sselement{2_{001}}{E}{0}$ element originating from the anisotropy blindness:
\begin{align}
&(1,2)_a  \xrightarrow{\sselement{4_{001}^{\pm}}{4_{001}^{+}}{\bf 0}} (2,1)_b ~,\\  \nonumber
&(1,2)_b  \xrightarrow{\sselement{4_{001}^{\pm}}{4_{001}^{-}}{\bf 0}} (2,1)_a ~,
\end{align}
and the requirements can be satisfied simultaneously by $\sselement{4_{001}^{+}}{m_{100}}{1/2,1/2,0}$ (or by $4_{001}^{-}$).  The same holds for the other bonds in the set $J_1$ and for $J_2$. 
In addition the spin rotation $4_{001}^{+}$ preserves the magnetic order (ferromagnetic moment along the $[001]$ direction), and therefore all the conditions are satisfied such that the LSWT spectrum of  $J_1 + J_2$ hosts this extra symmetry.
Lastly we can quickly check how the glide operates on bond $J_{3a}$:
\begin{align}
(1,1')_x  \xrightarrow{\{m_{100}|1/2,1/2,0\}} (2',2)_x ~,
\end{align}
but $(2',2)_x $ belongs to $J_{3b}$, so they are totally independent couplings and therefore no spin rotations can satisfy these requirements.
Therefore the model $J_1 + J_2 + J_3$ does not have this extra symmetry and the LSWT description correctly reflects the lattice symmetries.

Turning now to the effect of the enhanced symmetries on the magnon spectrum, one finds that the $J_1 + J_2$ model has a degeneracy on the line $[R\,X] = (0 ~1/2 ~u)$ in the Brillouin Zone.
Since the line is on the boundary of the Brillouin Zone and there are non-symmorphic elements we may need to consider projective representations of the relevant spin-space group $\mathcal{G}_{\rm LSWT}$ to explain the degeneracy. In this process one can show that the new glide element $\sselement{4_{001}^{+}}{m_{100}}{1/2,1/2,0}$ described above is responsible for the degeneracy.
We briefly outline here the theory of projective representations and the chain of reasoning for this specific case. For general discussion of projective representations and application to spin-space group see \cite{bradley2009mathematical, corticelli2022}.

{\it Projective representations.}  
A representation is said to be projective when $\Delta(h_i) \Delta(h_j) = \mu(h_i, h_j) \Delta(h_k)$, where $\Delta$ are matrix representations of little group elements $h_i \in \mathcal{G}^{\bs{k}}$ and $\mu(h_i, h_j) = \exp( - \mathrm{i}\bs{g}_i \cdot \bs{w}_j)$ is an element of the factor system, with $\bs{g}_i = h_i^{-1} \bs{k} - \bs{k}$ and $\bs{w}_j$ the translation associated to $h_j$.
If $\mu(h_i, h_j) = 1$ for all cases then we reduce to ordinary (non-projective) representations. 
If this is not the case we proceed by studying the representations of the central extension of the little co-group $\bar{\mathcal{G}}^{\bs{k}^*} = \bar{\mathcal{G}}^{\bs{k}} \otimes {Z}_g$ with kernel ${Z}_g$, the cyclic group of integers $0, 1, ..., (g-1)$.  The number $g$ comes from the parametrization of the factor system as $\mu(h_i, h_j) = \exp( 2 \pi \mathrm{i} a(h_i, h_j) / g)$, where $a(h_i, h_j) = 0,1,..., (g-1)$ and the group elements are of the kind $(h_i, \alpha)$ with product rule $(h_i, \alpha) (h_j, \beta) = (h_i h_j, \alpha + \beta + a(h_i, h_j))$.
Of all the irreps of $\bar{\mathcal{G}}^{\bs{k}^*}$ we are interested only in the ones giving the right factor system, that is the ones with $\Delta(E,\alpha) =  \exp( 2 \pi \mathrm{i} \alpha / g) \, \mathbb{I}$. Since the set of elements $(h_i, 0)$ is isomorphic to $\bar{\mathcal{G}}^{\bs{k}}$, we can now extrapolate the character tables of those irreps and build the table of projective irreducible representations of $\bar{\mathcal{G}}^{\bs{k}}$ (and therefore the one of $\mathcal{G}^{\bs{k}}$, adding the right phase factors coming from translations).

Turning now to the specific case of line $[R\,X] = (0 ~1/2 ~u)$ the little group is:
\begin{align}
\mathcal{G}^{[R\,X]}_{LSWT} = \sselement{E / 2_{001}}{E}{0}, \sselement{E / 2_{001}}{2_{001}}{0}, ~ \sselement{4_{001}^{\pm}}{m_{100}}{1/2,1/2,0}, ~ \sselement{4_{001}^{\pm}}{m_{010}}{1/2,1/2,0}
\end{align}
The factor system is not trivial since for example $\mu(\sselement{E}{2_{001}}{0}, \sselement{4_{001}^{\pm}}{m_{100}}{1/2,1/2,0}) =-1$. We then find the central extension group $\bar{\mathcal{G}}_{LSWT}^{[R\,X]^*}$ with $g =2$ as a group with $16$ elements:
\begin{align}
&\bar{\mathcal{G}}_{LSWT}^{[R\,X]^*} = ( G_{1} + G_{1} \times ([2_{001}||E],0) ) \times ( E + ([4_{010}^{+} || m_{100}], 0) ) \\
&G_1 = ([E||E],0), ~ ([E||E],1), ~ ([E||2_{001}],0), ~ ([E||2_{001}],1) \cong D_{2}
\end{align}
The irreducible representations can be obtained by conjugating the ones of the subgroup $D_{2}$ (1D) by the symmetry $([2_{001}||E],0)$ (trivial self-conjugation) and later by $([4_{010}^{+} || m_{100}], 0)$, which eventually pair some of them in two-dimensional representations.
Of these representations we are only interested in the ones with $\Delta( [E||E],1) =  -\, \mathbb{I}$ which give the character table for the little group $\mathcal{G}_{LSWT}^{[R\,X]}$ as in Tab.~\ref{tab:ProjReptableRX}.
Finally the symmetry transformations on the transverse spin components on the lattice select as representation:
\begin{align}
\label{eq:RXRep}
\rho^{[R\,X]}_{S_{\pm}} = \Gamma_{34}^{-}(2)
\end{align}
which is two-dimensional. Therefore we will observe a spurious two-dimensional nodal line in the LSWT spectrum, which it is not present taking in consideration non-linear terms or further interaction shells.
\begin{table*}[!htb]
\centering
\hbox to \linewidth{ \hss
\begin{tabular}{ |c|c|c|c|c|c|c|  }
 \hline
 $\mathcal{G}_{LSWT}^{[R\,X]}$~ & ~$\sselement{E}{E}{0}$~ & ~$\sselement{E}{2_{001}}{0}$~ & ~$\sselement{2_{001}}{E}{0}$
 ~ & ~$\sselement{4_{001}^{+}}{m_{100}}{1/2,1/2,0}$ ~ & ~$...$\\ 
 \hline
  $\Gamma_{34}^{-}$   & 2   & 0    &   -2   &  0    &  $...$\\
 \hline
 $\Gamma_{1a}^{-}$   & 1   & 1     &  -1  &  $\xi$   &  $...$ \\
 \hline
 $\Gamma_{1b}^{-}$    & 1   & 1    &  -1   &  -$\xi$   &  $...$ \\
 \hline
 $\Gamma_{2a}^{-}$   & 1    & -1   &  -1  &   $\xi$    &  $...$ \\
  \hline
 $\Gamma_{2b}^{-}$   & 1    & -1   &  -1  &   -$\xi$    &  $...$ \\
 \hline
\end{tabular}
\hss}
\caption{
Character table giving the irreducible representations of $\mathcal{G}_{LSWT}^{[R\,X]}$ with negative character for global spin rotation $\sselement{2_{001}}{E}{0}$, relevant for generic transverse spin components.
The phase factor is $\xi = \exp(\mathrm{i} \, (0 ~1/2 ~u) \cdot (\textstyle \frac{1}{2} \frac{1}{2} 0) ) = \mathrm{i}$ and the dot $(...)$ indicates the other redundant symmetries to complete the table.
}
\label{tab:ProjReptableRX}
\end{table*}
%

\end{widetext}

\bibliography{references}